\documentclass[12pt]{article}
\pdfoutput=1

\usepackage{putex}
\usepackage{graphicx}
\usepackage{caption}
\usepackage{amsmath}
\usepackage{array}
\usepackage{subcaption}
\usepackage{epstopdf}
\usepackage{enumerate}
\usepackage{cite}
\usepackage{youngtab}
\usepackage{tensor}
\usepackage{slashed}
\usepackage[aligntableaux=center]{ytableau}
\usepackage[utf8]{inputenc}
\usepackage{rotating}
\usepackage[
      colorlinks=true,
      linkcolor=blue,
      urlcolor=blue,
      filecolor=black,
      citecolor=red,
      ]{hyperref}

\newcommand{\abs}[1]{\left\lvert #1 \right\rvert}

\newcommand {\be} {\begin {equation}}
\newcommand {\ee} {\end {equation}}

\newcommand {\bes} {\begin {equation*}}
\newcommand {\ees} {\end {equation*}}

\newcommand{\es}[2] {\begin{equation} \label{#1} \begin{split} #2 \end{split} \end{equation}}

\newcommand{\Z}{\mathbb{Z}}

\newcommand{\C}{\mathbb{C}}

\newcommand{\cA}{{\mathcal A}}
\newcommand{\cB}{{\mathcal B}}
\newcommand{\cC}{{\mathcal C}}

\newcommand{\cN}{{\mathcal N}}
\newcommand{\cO}{{\mathcal O}}
\newcommand{\cP}{{\mathcal P}}

\newcommand{\cR}{{\mathcal R}}
\newcommand{\cS}{{\mathcal S}}
\newcommand{\cW}{{\mathcal W}}

\newcommand{\veps}{\varepsilon}

\newcommand{\beq}{\begin{equation}}
\newcommand{\eeq}{\end{equation}}

\def\ie{\begin{equation}\begin{aligned}}
\def\fe{\end{aligned}\end{equation}}

\newcommand{\m}{\mu}

\newcommand{\mf}{\mathfrak }

\numberwithin{equation}{section}

\def\<{\langle}
\def\>{\rangle}

\begin{document}

\preprint{PUPT-2570}

\institution{PU}{Joseph Henry Laboratories, Princeton University, Princeton, NJ 08544, USA}

\title{
Absence of $D^4 R^4$ in M-Theory From ABJM
}

\authors{Damon J.~Binder, Shai M.~Chester, and Silviu S.~Pufu}

\abstract{
Supersymmetry allows a $D^4 R^4$ interaction in M-theory, but such an interaction is inconsistent with string theory dualities and so is known to be absent.   We provide a novel proof of the absence of the $D^4 R^4$ M-theory interaction by calculating 4-point scattering amplitudes of 11d supergravitons from ABJM theory.  This calculation extends a previous calculation performed to the order corresponding to the $R^4$ interaction.  The new ingredient in this extension is the interpretation of the fourth derivative of the mass deformed $S^3$ partition function of ABJM theory, which can be determined using supersymmetric localization, as a constraint on the Mellin amplitude associated with the stress tensor multiplet 4-point function.  As part of this computation, we relate the 4-point function of the superconformal primary of the stress tensor multiplet of any 3d ${\cal N} = 8$ SCFT to some of the 4-point functions of its superconformal descendants. We also provide a concise formula for a general integrated 4-point function on $S^d$ for any $d$.
}
\date{}

\maketitle

\tableofcontents

\section{Introduction and Summary}

M-theory is a rather mysterious ultraviolet (UV) completion of eleven-dimensional supergravity \cite{Witten:1995ex}.  It describes the dynamics of massless gravitons and their superpartners,\footnote{It is possible that M-theory also contains trans-Planckian particles.} and as such its main observables are the scattering amplitudes of these massless particles.  For general momenta, supersymmetry requires that the 4-graviton S-matrix take a factorized form ${\cal A}(\eta_i,s, t) = {\cal A}_{\rm SG, tree}(\eta_i, s, t) f(s, t)$ \cite{Wang:2015aua}. The first factor is the tree-level scattering amplitude computed in 11d supergravity and depends on the polarizations $\eta_i$ of the four gravitons as well as the Mandelstam invariants.  The second factor is an arbitrary symmetric function $f$ of the Mandelstam invariants $s$, $t$, and $u = -s - t$.  In the small momentum expansion (or equivalently in the expansion in the 11d Planck length $\ell_p$), 11d supersymmetry allows the following terms in $f$:
 \es{A11D}{
  f(s, t) =  1 + \ell_{p}^6 f_{R^4}(s, t) 
   + \ell_{p}^9 f_\text{1-loop}(s, t) + \ell_{p}^{10} f_{D^4 R^4}(s, t)  + \ell_{p}^{12} f_{D^6 R^4}(s, t) 
   + \cdots \, 
 }
Terms multiplying $\ell_p^n$ are homogeneous of degree $n$ in $s$ and $t$, as required on dimensional grounds. The functions of $s$ and $t$ that constitute the coefficients of $\ell_p^n$ are given suggestive names:  $f_{D^{2m} R^4}$ are symmetric polynomials in $s$, $t$, $u$ that represent the contributions of the contact Feynman diagrams with $D^{2m} R^4$ as well as tree-level exchange diagrams that have the same momentum scaling;  $f_{\text{1-loop}}$ is the one-loop supergravity Feynman diagram, etc.  All the loop corrections to supergravity can be computed in principle from the 11d supergravity Lagrangian. On the other hand only the lowest few protected $f_{D^{2m} R^4}$ corrections to supergravity can be determined by relating them to type II perturbative string theory computations and non-renormalization theorems.\footnote{If M-theory is compactified to 4d, one obtains 4d ${\cal N} = 8$ SUGRA theory supplemented by higher derivative corrections \cite{Cremmer:1978ds,Cremmer:1979up}.  Constraints on higher derivative counterterms in 4d ${\cal N} = 8$ SUGRA have been studied in \cite{Elvang:2010jv,Beisert:2010jx,Elvang:2010kc,Bossard:2011tq} (see also \cite{Elvang:2015rqa}).} These take the form \cite{Green:1997as, Russo:1997mk, Green:2005ba,Green:1998by, Pioline:2015yea}
 \es{fR4}{
  f_{R^4} (s, t) = \frac{ s t u}{3 \cdot 2^7}\,, ~~~~~ f_{D^4R^4}(s,t) =0\,, ~~~~~ f_{D^6R^4}(s,t) = {(stu)^2\over 15\cdot 2^{15}}\,,
 }
with $u=-s-t$ as above. The $D^4R^4$ contribution is absent, but would otherwise be consistent with 11d supersymmetry. The goal of this paper is to derive the vanishing of $f_{D^4 R^4}$ purely from 3d CFT using AdS/CFT.

It was proposed in \cite{Chester:2018aca} following earlier work \cite{Heemskerk:2009pn,Fitzpatrick:2011ia,Penedones:2010ue} that an alternative way of determining the 11d 4-graviton S-matrix is from the flat space limit of stress tensor multiplet 4-point correlation functions in the superconformal field theory (SCFT) on $N$ coincident M2-branes.  This theory is part of a family of $U(N)_k \times U(N)_{-k}$ gauge theories coupled to bifundamental matter whose Lagrangian descriptions are due to Aharony, Bergman, Jafferis, and Maldacena (ABJM) \cite{Aharony:2008ug}.  For general $N$ and $k$, ABJM theory is dual to the $AdS_4 \times S^7 / \Z_k$ background of M-theory and is the effective theory on $N$ coincident M2-branes placed at a $\C^4 / \Z_k$ singularity in the transverse space.  We will only focus on the cases $k=1$ or $2$ where supersymmetry is enhanced to ${\cal N} = 8$ \cite{Aharony:2008ug,Bashkirov:2010kz,Gustavsson:2009pm,Kwon:2009ar,Benna:2009xd} from the ${\cal N} = 6$ manifestly preserved at all $k$ \cite{Aharony:2008ug}.  Instead of parameterizing these theories by $N$, we will find it convenient to use the quantity $c_T \sim N^{3/2}$, the coefficient of the canonically-normalized stress-tensor two point function, which has been calculated to all orders in $1/N$ through supersymmetric localization \cite{Chester:2014fya} using the results of \cite{Jafferis:2010un} and \cite{Closset:2012vg}.

More concretely, one can consider the four-point function $\langle SSSS\rangle$ of the scalar bottom component $S$ of the $\cN = 8$ stress tensor multiplet. It was shown in \cite{Chester:2018aca} that the ${\cal N} = 8$ superconformal Ward identity for the Mellin transform \cite{Fitzpatrick:2011ia,Penedones:2010ue} $M_\text{tree}^{SSSS}$ of the tree level $\langle SSSS\rangle$ correlator has a finite number of solutions at every order in the $1/c_T$ expansion:
\es{MellinIntro}{
M^{SSSS}_\text{tree}(s,t)=&c_T^{-1} B_1^1M_\cS^{1}+c_T^{-\frac53}\left[B_4^4 M_\cS^{4}+B_1^4M_\cS^{1} \right]+c_T^{-\frac{19}{9}}\left[B_6^6 M_\cS^{6}+B_4^6M_\cS^{4}+B_1^6M_\cS^{1} \right]\\
&+c_T^{-\frac{7}{3}}\left[ B_7^7M_\cS^{7}+B_6^7M_\cS^{6}+B_4^7M_\cS^{4}+B_1^7M_\cS^{1} \right]+\dots\,,
}
where $s,t$ are Mellin space variables that are related to the 11d Mandelstam variables in the flat space limit, $M_\cS^p$ are functions of $s, t$ that grow as the $p$th power at large $s, t$, and the $B$'s are numerical coefficients unfixed by 3d supersymmetry. If one can determine these $B$'s up to order $1/c_T^n$, then by taking the flat space limit \cite{Heemskerk:2009pn} one can reproduce the 4-graviton scattering amplitude and read off the function $f$ to order $\ell_p^{9(n-1)}$.  This procedure was carried out in  \cite{Chester:2018aca} to first non-trivial order:  by computing two distinct CFT quantities as a function of $N$ (namely $c_T$ as well as a $1/2$-BPS OPE coefficient), Ref.~\cite{Chester:2018aca} was able to fix both $B_i^4$ and thereby reproduce exactly the known value of $f_{R^4}(s, t)$.  (See also~\cite{Chester:2018dga} for an analogous computation in 6d.)  The goal of the present paper is to use ABJM theory to find an additional constraint such that all three $B^6_i$ can be fixed to zero, which implies that $f_{D^4 R^4} = 0$.

Developing the momentum expansion up to $D^4 R^4$ order presents new challenges compared to the computation up to $R^4$ order performed in \cite{Chester:2018aca}.  In \cite{Chester:2018aca}, the computation of the additional CFT quantity besides $c_T$ that was needed involved a trick based on the fact that, as any ${\cal N} \geq 4$ 3d SCFT, ABJM theory contains a one-dimensional topological sector \cite{Chester:2014mea,Beem:2016cbd}.  Such a trick does not seem easily generalizable to the computation of other quantities.  Nevertheless, one may hope to to beyond $R^4$ order because in 11d all terms up to $D^6 R^4$ (thus including $D^4 R^4$) preserve some amount of supersymmetry, and thus one may hope to be able to use supersymmetric localization to compute just enough quantities in the field theory in order to recover all these terms in 11d.

In ABJM theory, quite a few BPS quantities can be computed using supersymmetric localization using \cite{Kapustin:2009kz,Jafferis:2010un}.  To go to order $D^4 R^4$, it is enough to consider the free energy on a round three-sphere in the presence of real mass deformations.  When ABJM theory is viewed as an ${\cal N} = 2$ SCFT, it has $SU(4)$ flavor symmetry, and, since the Cartan of $SU(4)$ is three-dimensional, it also admits a three-parameter family of real mass deformations. (See for instance \cite{Freedman:2013oja} where these deformations were studied at leading order $1/N$.)  The $S^3$ free energy $F$ in the presence of two such mass parameters $m_1$ and $m_2$ was computed to all orders in $1/N$ in \cite{Nosaka:2015iiw} using the Fermi gas formalism developed in \cite{Marino:2011eh}.  To make connection with the four-point function of the stress tensor multiplet, we consider the fourth derivatives\footnote{All other fourth derivatives vanish or are linearly dependent on the two mentioned in the main text.} $\frac{\partial^4 F}{\partial m_1^4}\big|_{m_1 = m_2 =0}$ and $\frac{\partial^4 F}{\partial m_1^2 \partial m_2^2}\big|_{m_1 = m_2 =0}$, which, by analogy with the analysis of \cite{Closset:2012vg} for two-point functions, can be related to integrated four-point correlators in the SCFT\@.  Calculating these integrated correlators using the solution to the Ward identity at order $D^4 R^4$ and comparing with the fourth derivatives of $F$ mentioned above, one can fix all three $B^6_i=0$ so that $f_{D^4 R^4}(s, t) = 0$.

It is worth pointing out that part of the difficulty in performing this computation is that all previous studies \cite{Chester:2014mea,Agmon:2017xes,Chester:2014fya,Chester:2018aca,Chester:2018lbz} in 3d ${\cal N} = 8$ SCFTs focused on the four-point function of the superconformal primary of the stress tensor multiplet, which is the scalar operator $S$ mentioned above of scaling dimension $\Delta_S = 1$ transforming in the ${\bf 35}_c$ of the $SO(8)_R$ $R$-symmetry.  However, the fourth mass derivatives of the $F$ are more directly related to integrated four-point functions of a linear combination of $S$ and another operator $P$ that belongs to the same superconformal multiplet as $S$.  The operator $P$ is a pseudoscalar of scaling dimension $\Delta_P = 2$ transforming in the ${\bf 35}_s$ of $SO(8)_R$.  As part of our computation, we will therefore derive expressions for the four-point functions $\langle S S PP \rangle$ and $ \langle PPPP \rangle$ in terms of the more easily computable $\langle SSSS \rangle$.

Another notable feature of our computation is a concise expression for integrals over $S^d$ of 4-point functions of scalar operators in CFTs. In particular, we find that the integral over $4d$ variables reduces to an integral over the two conformally-invariant cross ratios $U$ and $V$ of the 4-point function multiplied by a $\overline{D}(U,V)$ function, which naturally shows up in tree level calculations in AdS$_{d+1}$ \cite{Rastelli:2017udc}. (See Eqs.~\eqref{4pointS3Int} and~\eqref{IRewrite2}.) While in this work we only apply this result to $d=3$ and the specific operators we are interested, this formula applies to any CFT with or without supersymmetry. 

The rest of this paper is organized as follows.  We start in Section~\ref{4points} with a brief review of the relevant four-point functions in ${\cal N} = 8$ SCFTs and derive the relations between them that were mentioned in the previous paragraph.  In Section~\ref{INTEGRATED}, we discuss the relation between the fourth mass derivatives of the $S^3$ free energy and integrated correlation functions in the SCFT\@.   In Section~\ref{ABJM} we apply these results to ABJM theory, and by taking the flat space limit of the SCFT correlators show that $f_{D^4 R^4}$ vanishes, as expected.  Lastly, we end with a brief discussion of our results in Section~\ref{DISCUSSION}.  Various technical details are relegated to the Appendices as well as to an auxiliary \verb|Mathematica| file included with this arXiv submission, containing the Ward identities and the large $c_T$ expressions for the Mellin amplitudes and position space correlators that we computed. 

\section{Stress tensor multiplet 4-point functions}
\label{4points}

Let us begin by discussing the structure of the stress energy tensor multiplet in $\cN = 8$ SCFTs and then derive the relation between the 4-point function of the ${\bf 35}_c$ scalar $S$ and the 4-point functions involving the ${\bf 35}_s$ pseudoscalar $P$ that were mentioned in the Introduction.  

\subsection{The ${\cal N} = 8$ stress tensor multiplet}
In addition to the scaling dimension $1$ scalar operator $S$ and the dimension $2$ pseudoscalar operator $P$ transforming in the ${\bf 35}_c$ and ${\bf 35}_s$ irreps of the $SO(8)_R$ $R$-symmetry, the ${\cal N} = 8$ stress tensor multiplet also contains a fermionic operator $\chi_\alpha$ of dimension $3/2$ in the ${\bf 56}_v$, the $R$-symmetry current $j_\mu$ in the ${\bf 28}$, the supercurrent $\psi_{\mu \alpha}$ in the ${\bf 8}_v$, and the stress tensor itself, $T_{\mu\nu}$---See Table~\ref{stressTable}. 

\begin{table}
\begin{center}
\begin{tabular}{c|c|c|c}
  Operator & $\Delta$ & Spin & $\mathfrak{so}(8)$ charge \\
  \hline
  $S$ & 1 & 0 & ${\bf 35}_c$ \\
  $\chi$ & 3/2 & 1/2 & ${\bf 56}_v$ \\
  $P$ & 2 & 0 & ${\bf 35}_s$ \\
  $j$& 2 & 1 & ${\bf 28}$ \\
  $\psi$ & 5/2 & 3/2 & ${\bf 8}_v$ \\
  $T$ & 3 & 2 & ${\bf 1}$
\end{tabular}
\caption{Operators in the $\cN = 8$ stress energy tensor multiplet.}
\label{stressTable}
\end{center}
\end{table}

In order to write down the supersymmetry variations relating all these operators, we need an efficient way to keep track of the various $SO(8)$ representations that appear.  All representations of $SO(8)$ can be produced by symmetrizing and antisymmetrizing the ${\bf 8}_c$, ${\bf 8}_s$, and ${\bf 8}_v$ representations. We use the indices $I, J, ...$ for ${\bf 8}_c$; $A, B, ... $ for ${\bf 8}_s$; and $a, b, ...$ for ${\bf 8}_v$.\footnote{These three representations are all equivalent due to $SO(8)$ triality, and so while we can think of the ${\bf 8}_v$ as the vector representation while ${\bf 8}_c$ and ${\bf 8}_s$ are the two (real, inequivalent) spinor representations, this assignment is arbitrary.}  All three representations are real and so indices are raised and lowered using the Kronecker delta symbol. Equivalently, the Kronecker delta symbol can be used to form a singlet from ${\bf 8}_i \otimes {\bf 8}_i$, an operation we denote by $\cdot$ in index free notation (e.g.~$v \cdot w \equiv v_A w^A$ if $v$ and $w$ transform in the ${\bf 8}_s$). In addition to the Kronecker delta symbol, $SO(8)$ admits another invariant tensor $E_{aIA}$ (see Appendix~\ref{EMatrices} for explicit expressions), which can be used to produce the ${\bf 8}_k$ from ${\bf 8}_i\otimes{\bf 8}_j$ whenever $i$, $j$, and $k$ are distinct.  In index free notation, we can represent this as a wedge product $\wedge:{\bf 8}_i\otimes{\bf 8}_j\rightarrow{\bf 8}_k$ (e.g.~$(v \wedge w)_a = E_{a I A} v^I w^A$ if $v$ and $w$ transform in the ${\bf 8}_c$ and ${\bf 8}_s$, respectively).\footnote{\label{Clifford}The $E_{a I A}$ can be thought of as chiral $SO(8)$ gamma matrices.  The Clifford algebra implies $$(X_1\wedge Y_1)\cdot (X_2\wedge Y_2) + (X_1\wedge Y_2)\cdot (X_2\wedge Y_1) = 2(X_1 \cdot X_2)(Y_{1} \cdot Y_2) \,. $$  By manipulating the Clifford algebra, one can derive other useful relations, for instance $$(X\wedge Y_1)\wedge Y_2 + (X\wedge Y_2)\wedge Y_1 = 0\,.$$}

Then, we can represent the operators of the stress tensor multiplet as $S_{IJ}$, $P_{AB}$, $\chi_\alpha^{AI}$, $j_\mu^{IJ}$ (or $j_\mu^{AB}$ or $j_\mu^{ab}$), $\psi_{\mu\alpha}^a$, $T_{\mu\nu}$.  The scalars $S$ and $P$ are rank-two traceless symmetric tensors of ${\bf 8}_c$ and ${\bf 8}_s$, respectively, $j_\mu$ is an anti-symmetric tensor of any eight-dimensional representation, and $\chi$ obeys $\chi_{AI} E^{AIa} = 0$ in order to select the ${\bf 56}_v$ representation from the product ${\bf 8}_s \otimes {\bf 8}_c = {\bf 56}_v \oplus {\bf 8}_v$.  Including all the $SO(8)_R$ indices quickly becomes unwieldy, so instead we will use polarization vectors:  
we will denote vectors in ${\bf 8}_c$ by $Y$, those in ${\bf 8}_s$ by $X$, and those in ${\bf 8}_v$ by $Z$. Then we can define the operators
 \es{Polarizations}{ 
  S(\vec{x},Y) &= S_{IJ}(\vec{x})Y^IY^J \,, \qquad \ \ \ \ \ 
 P(\vec{x},X) = P_{AB}(\vec{x})X^AX^B \,, \\
  \chi_\alpha(\vec{x}, X, Y) &= \chi_\alpha^{AI}(\vec{x}) X_A Y_I \,, \qquad 
  j_\mu(\vec{x}, Y_1, Y_2) = j_\mu^{IJ}(\vec{x}) Y_{1I} Y_{2J} \,, \\
   \psi_{\mu\alpha}(\vec{x}, Z) &= \psi_{\mu\alpha}^a(\vec{x}) Z_a \,.
 }
To implement the tracelessness of $S_{IJ}$ and $P_{AB}$, we demand that $Y\cdot Y\equiv Y^IY_I = 0$ in the definition of $S(\vec{x}, Y)$ and similarly for $X$ in the definition of $P(\vec{x}, X)$.\footnote{Strictly speaking we should view $S$ and $P$ as functions of two distinct auxiliary fields; for instance, $S(\vec{x},Y_1,Y_2) = S_{IJ}(\vec{x})Y_1^IY_2^J$, subject to the conditions that $S(\vec{x},Y_1,Y_2) = S(\vec{x},Y_2,Y_1)$ and that $Y_1\cdot Y_2 = 0$. But we can always uniquely reproduce the full $SO(8)$ structures by restricting to $Y_1 = Y_2$ and it is usually convenient to do so.}  Likewise, to implement the condition $E_{AI}^a\chi^{AI}(\vec{x}) = 0$, we require $X \wedge Y = 0$.   We can automatically satisfy this condition by choosing $X = Y\wedge Z$ for some $Z\in{\bf 8}_v$; this is now a redundant parametrization as there exists a (unique up to normalization) vector $Z_Y$ for which $Y\wedge Z_Y = 0$.  Lastly, since the $R$-symmetry current $j^\m$ transforms in the adjoint representation ${\bf 28} \in( {\bf 8}_c\otimes{\bf 8}_c)_a$ we can polarize it with two vectors $Y_1$ and $Y_2$, but all expression must be antisymmetric in these two vectors. Alternatively we could polarize it with $X_1$ and $X_2$ or $Z_1$ and $Z_2$, depending on whichever is most convenient.

After this long introduction on notation, we can write down how Poincar\'e supersymmetry generated by the supercharges $Q^{\alpha a}$ relate the operators in the stress tensor multiplet:
 \es{SUSYvars}{
\delta^\alpha(Z)S(\vec{x},Y) &= \chi^\alpha(\vec{x},Z\wedge Y,Y) \,, \\
\delta^\alpha(Z)\chi^\beta(\vec{x},X,Y) &= \frac {\epsilon^{\alpha\beta}} {\sqrt 2}P(\vec{x},X,Y\wedge Z)+ \frac {1} 2 \sigma^{\alpha\beta}_\mu j^\mu(\vec{x},X,Y\wedge Z) + i\sigma^{\alpha\beta}_\mu\partial^\mu S(\vec{x},X\wedge Z,Y) \,, \\
\delta^\alpha(Z)P(\vec{x},X) &= i\sqrt 2\sigma_\mu^{\alpha\beta}\partial^\mu\chi_\beta(\vec x,X,X\wedge Z)\,, \\
&\text{etc.}
}
Here, $\delta^\alpha(Z)$ represents the action of $Q^{\alpha a} Z_a$ on the various operators and $\sigma_\mu$ are the 3d gamma matrices, which can be taken to be just Pauli matrices.  The supersymmetry variations of $j^\mu$, $\psi^{mu\alpha}$, and $T^{\mu\nu}$ that were omitted from \eqref{SUSYvars} will not be needed in this work.

\subsection{Ward identities}

To derive the relations superconformal symmetry imposes between the four-point functions of the stress tensor multiplet operators, it is enough to first determine the most general form of these four-point functions that is consistent with conformal symmetry, and then require that they be invariant under the Poincar\'e SUSY transformations in \eqref{SUSYvars}.  (One does not gain any additional information by also imposing invariance under the superconformal generators $S^{\alpha a}$ because invariance under $S^{\alpha a}$ is guaranteed by invariance under $Q^{\alpha a}$ and under the special conformal generators $K^\mu$.)  For example, conformal symmetry implies that the $\langle SSSS \rangle$ and $\langle PPPP \rangle$ correlators take the form
 \es{SPFourPoint}{
  \langle S(\vec{x}_1, Y_1) \cdots S(\vec{x}_4, Y_4) \rangle &= \frac{1}{x_{12}^2 x_{34}^2} \biggl[ \cS_{1}(U, V) Y_{12}^2 Y_{34}^2 
   + \cS_{2}(U, V) Y_{13}^2 Y_{24}^2  + \cS_{3}(U, V) Y_{14}^2 Y_{23}^2 \\
    &\hspace{-1in}{}+\cS_{4}(U, V) Y_{13} Y_{14} Y_{23} Y_{24} +  \cS_{5}(U, V) Y_{12} Y_{14} Y_{23} Y_{34} 
     + \cS_{6}(U, V) Y_{12} Y_{13} Y_{24} Y_{34}  
   \biggr] \,, \\
  \langle P(\vec{x}_1, X_1) \cdots P(\vec{x}_4, X_4) \rangle &= \frac{1}{x_{12}^4 x_{34}^4} \biggl[ \cP_{1}(U, V) X_{12}^2 X_{34}^2 
   + \cP_{2}(U, V) X_{13}^2 X_{24}^2  + \cP_{3}(U, V) X_{14}^2 X_{23}^2 \\
    &\hspace{-1in}{}+\cP_{4}(U, V) X_{13} X_{14} X_{23} X_{24} +  \cP_{5}(U, V) X_{12} X_{14} X_{23} X_{34} 
     + \cP_{6}(U, V) X_{12} X_{13} X_{24} X_{34}  
   \biggr] \,, 
 }
where the $\cS_i$ and the $\cP_i$ are functions of the conformal cross-ratios 
 \es{UVDef}{
  U \equiv \frac{x_{12}^2 x_{34}^2}{x_{13}^2 x_{24}^2} \,, \qquad 
   V \equiv \frac{x_{14}^2 x_{23}^2}{x_{13}^2 x_{24}^2} \,. 
 }
Note that not all the functions $\cS_i$ and $\cP_i$ are independent.  Crossing symmetry implies the relations 
 \es{Crossing}{
\cS_2(U,V) &= U\cS_1\left(\frac 1 U,\frac V U\right)\,, \qquad \ \   \cS_3(U,V) = \frac U V\cS_1(V,U)  \,, \\
\cS_5(U,V) &= U\cS_4\left(\frac 1 U,\frac V U\right)\,, \qquad \ \ \cS_6(U,V) = \frac U V\cS_4(V,U) \,, \\
\cP_2(U,V) &= U^2\cP_1\left(\frac 1 U,\frac V U\right)\,, \qquad  \cP_3(U,V) = \frac {U^2}{V^2}\cP_1(V,U) \,, \\
\cP_5(U,V) &= U^2\cP_4\left(\frac 1 U,\frac V U\right)\,, \qquad \cP_6(U,V) = \frac {U^2}{V^2}\cP_4(V,U) \,.
 }
Likewise, $\langle SSPP \rangle$ takes the form
 \es{SSPP}{
  &   \langle S(\vec{x}_1, Y_1) S(\vec{x}_2, Y_2) P(\vec{x}_3, Y_3) P(\vec{x}_4, Y_4) \rangle = \\
     &\qquad\frac{1}{x_{12}^2 x_{34}^4} \biggl[ \cR_1 Y_{12}^2 X_{34}^2 
   + \cR_{2} \left[ (Y_1 \circ Y_2) \cdot (X_3 \circ X_4) \right]^2  +\cR_{3} (Y_1 \circ Y_2) \cdot (X_3 \circ X_4) Y_{12} X_{34}
   \biggr],
 } 
where the $\cR_i$ are also functions of $U$ and $V$ and we used the product $\circ : {\bf 8}_i\otimes{\bf 8}_i \rightarrow \bf 28$, normalized such that $(Y_1\circ Y_2)\cdot(Y_3\circ Y_4) = \frac 1 4 (Y_{13}Y_{24}-Y_{14}Y_{23}) $.\footnote{Using the identities in Footnote~\ref{Clifford}, we can derive $(Y_1\circ Y_2)\cdot(X_3\circ X_4) = \frac 12 \left[ (X_3\wedge Y_1)\cdot (X_4\wedge Y_2) - (X_3\wedge Y_2)\cdot (X_4\wedge Y_1)\right]$.}  For other correlation functions (which are not needed in the rest of this paper), see Appendix~\ref{WardAppendix}.

For the application presented in this paper, we only need to express the four-point functions $\langle SSPP \rangle$ and $\langle PPPP \rangle$ in terms of $\langle SSSS \rangle$.  These relations can be determined by substituting the general form of the four-point functions into the identities
 \es{IdentityList}{
  \delta \langle SSS\chi \rangle &= 0 \,, \\
  \delta \langle SSP\chi \rangle &= 0 \,, \\
  \delta \langle SPP\chi \rangle &= 0 \,, \\
  \delta \langle PPP\chi \rangle &= 0 \,.
 }
In particular, from the first equation in \eqref{IdentityList}, we determine $\langle SS\chi\chi \rangle$ and $\langle SSSj \rangle$ in terms of $\langle SSSS \rangle$, as well as relations on $\langle SSSS \rangle$.  Then, from the second line of \eqref{IdentityList}, we determine $\langle SP\chi\chi \rangle$, $\langle SSPP \rangle$, and $\langle SSPj \rangle$.   Then, from the third line of \eqref{IdentityList}, we determine $\langle PP\chi\chi\rangle$ and $\langle SPPj\rangle$.  Lastly, from the fourth line of \eqref{IdentityList}, we determine $\langle PPPP \rangle$ and $\langle PPPj\rangle$.  See also Table~\ref{wardCorr}.

\begin{table}
\begin{center}
\begin{tabular}{|c|l l|l l l|}\hline
Variation & \multicolumn{2}{c|}{Correlators Used} & \multicolumn{3}{c|}{Correlators Obtained} \\ \hline
$\delta\langle SSS\chi\rangle$ & $\langle SSSS\rangle$       &                       & $\langle SS\chi\chi\rangle$ & $\langle SSSj\rangle$ & \\
$\delta\langle SSP\chi\rangle$ & $\langle SS\chi\chi\rangle$ &                       & $\langle SP\chi\chi\rangle$ & $\langle SSPP\rangle$ & $\langle SSPj\rangle$ \\
$\delta\langle SPP\chi\rangle$ & $\langle SP\chi\chi\rangle$ & $\langle SSPP\rangle$ & $\langle PP\chi\chi\rangle$ & $\langle SPPj\rangle$ & \\
$\delta\langle PPP\chi\rangle$ & $\langle PP\chi\chi\rangle$ &                       & $\langle PPPP\rangle$       & $\langle PPPj\rangle$ & \\ \hline
\end{tabular}
\end{center}
\caption{Taking supersymmetric variations to compute correlators. By setting the variation in the first column to zero, we can use the correlators in the second column to compute the correlators in the third column.}
\label{wardCorr}
\end{table}

In practice, plugging \eqref{SPFourPoint}, \eqref{SSPP}, and the analogous equations in Appendix~\ref{WardAppendix} into \eqref{IdentityList} is an onerous but straightforward task that can be greatly simplified using {\tt Mathematica}.  Our results are as follows.  From the first equation in \eqref{IdentityList}, we can show that the ${\cal S}_i$ obey the Ward identities
\begin{equation}\begin{split}
\partial_U \cS_4(U,V) &= \frac 1 U\cS_4(U,V) + \left(\frac 1 U -\partial_U-\partial_V\right)\cS_2(U,V)+\left(\frac 1 U + (U-1)\partial_U+V\partial_V\right)\cS_3(U,V)\,,\\ 
\partial_V \cS_4(U,V) &= -\frac 1 {2V}\cS_4(U,V) - \frac 1 V\left(1 - U\partial_U  + (1-U) \partial_V\right)\cS_2(U,V)-\left(\partial_U+\partial_V\right)\cS_3(U,V) \,,
\end{split}\end{equation}
along with other identities which can be derived using the crossing relations \eqref{Crossing}.  It can be checked that these equations are equivalent to the Ward identities obtained in \cite{Dolan:2004mu}. The expressions for the functions ${\cal R}_i(U, V)$, ${\cal S}_i(U, V)$, and ${\cal P}_i(U, V)$ that appear in \eqref{SSPP} and \eqref{SPFourPoint} in the $\langle SSPP \rangle$, $\langle SSSS \rangle$, and $\langle PPPP \rangle$ correlators, respectively, are related as
\es{wards}{
\mathcal{R}_i(U,V)=&\mathcal{D}^R_i(U,V,\partial_U,\partial_V) \cS_1(U,V)\,,\\
\mathcal{P}_i(U,V)=&\mathcal{D}^P_{ij}(U,V,\partial_U,\partial_V) \cS_i(U,V)\,,
}
where the differential operators $\mathcal{D}^R_i(U,V,\partial_U,\partial_V) $ and $\mathcal{D}^P_{ij}(U,V,\partial_U,\partial_V) $ are rational functions $U$ and $V$ and have at most 2 and 4 derivatives, respectively, and are given explicitly in Eqs.~\eqref{SSPP1}--\eqref{P4} in Appendix~\ref{WardAppendix}. The other correlators mentioned in Table~\ref{wardCorr} can also be written as differential operators acting on the $\cS_i$; their explicit expression can be found in the attached {\tt Mathematica} notebook.

\subsection{A check:  superconformal blocks}

A stringent check on the formulas \eqref{wards} as defined in \eqref{SSPP1}--\eqref{P4} is that they should map superconformal blocks to superconformal blocks.  In particular, if we take the ${\cal S}_i$ to correspond to a superconformal block, then these equations determine the corresponding superconformal blocks in the $\langle SSPP \rangle$ and $\langle PPPP \rangle$ correlators.  The fact that these equations produce a finite linear combinations of conformal blocks is nontrivial. 

As a simple example, the superconformal block corresponding to the $s$-channel exchange of the stress tensor multiplet in the $\langle SSSS \rangle$ correlator is \cite{Chester:2014fya}
 \es{StressBlock}{
  {\cal S}_i =  \begin{pmatrix}
   \frac14 \left(-g_{1, 0} + g_{3, 2}\right) & 0 & 0 &0 & g_{1, 0} + g_{2, 1} & g_{1, 0} - g_{2, 1} 
   \end{pmatrix}\,,
 }
where $g_{\Delta, \ell}$ are the conformal blocks written in the normalization used in \cite{Chester:2014fya}.  From \eqref{SSPP1} we find
 \es{StressBlockR}{
  {\cal R}_i = \begin{pmatrix}
   \frac 12 g_{3, 2} & 0 & -4 g_{2, 1} 
  \end{pmatrix} \,,
 }
and \eqref{P1}--\eqref{P4} along with their crossed versions imply
 \es{StressBlockP}{
  {\cal P}_i = \begin{pmatrix}
   g_{3, 2} & 0 & 0 & 0 & g_{2, 1} & -g_{3,2}
  \end{pmatrix} \,.
 }
Superconformal blocks for other multiplets can be worked out in a similar way.

\section{Integrated correlators on $S^3$}
\label{INTEGRATED}

Having described the four-point function of the scalar and pseudo-scalar operators in the stress tensor multiplet of an ${\cal N} = 8$ SCFT, let us now connect these quantities to the fourth derivatives of the $S^3$ partition function with respect to various mass parameters.  Before delving into the details of these mass deformations, let us note that the formulas \eqref{SPFourPoint}--\eqref{SSPP} also hold on a round $S^3$, with the only modification that the quantity $\vec{x}_{ij}$ should undergo the replacement
 \es{S3Distance}{
  \vec{x}_{ij} \to \frac{\vec{x}_{ij}}{\sqrt{1 + \frac{x_i^2}{4r^2}} \sqrt{1 + \frac{x_j^2}{4r^2}}}
   = \Omega(\vec{x}_i)^{1/2} \Omega(\vec{x}_j)^{1/2} \vec{x}_{ij}
 }
everywhere.  Here, $r$ is the radius of the three-sphere, and the three-sphere is taken to have the metric
 \es{S3Metric}{
  ds^2 = \Omega(\vec{x})^2 d\vec{x}^2 \,, \qquad \Omega(\vec{x}) \equiv \frac{1}{1 + \frac{x^2}{4r^2} } \,.
 }
The RHS of \eqref{S3Distance} is just the chordal distance between two points on $S^3$.  In particular, the replacement \eqref{S3Distance} leaves unchanged the conformally-invariant cross-ratios $U$ and $V$ defined in \eqref{UVDef}.

\subsection{Three-parameter family of real mass deformations}

We are interested in mass deformations on $S^3$ which preserve sufficient supersymmetry to compute the partition function using supersymmetric localization. This requires at minimum ${\cal N} = 2$--preserving mass deformations.  Viewed as an ${\cal N} = 2$ SCFT, any ${\cal N} = 8$ SCFT possesses an $\mf{su}(4)$ flavor symmetry generated by the subalgebra of $\mf{so}(8)_R$ which commutes with the $\cN = 2$ $R$-symmetry $\mf{u}(1)_R$.  In $\cN = 2$ SCFTs, real mass parameters are associated with conserved current multiplets, because they can be thought of as arising from giving supersymmetry-preserving expectation values to the scalars in the background vector multiplets that couple to the conserved current multiplets.  In particular, the operators of an $\cN = 2$ conserved current multiplet generating a symmetry algebra $\mf{g}$ with hermitian generators $T^a$ are: a $\Delta = 1$ scalar $J = J^a T^a$, a $\Delta = 2$ pseudo-scalar $K = K^a T^a$, and the conserved current $j_\mu = j_\mu^a T^a$.  If we normalize the flat space two-point functions at separated points as 
 \es{NormTwoPoint}{
  \langle j_\mu^a(\vec{x}) j_\nu^b(0) \rangle &= \frac{\tau \tr (T^a T^b)}{16\pi^2} \left( \delta_{\mu\nu} \partial^2 - \partial_\mu \partial_\nu \right) \frac{1}{x^2} \,, \\
   \langle J^a(\vec{x}) J^b(0) \rangle &= \frac{\tau \tr (T^a T^b)}{16 \pi^2 x^2}\,, \\
    \langle K^a(\vec{x}) K^b(0) \rangle &= \frac{\tau \tr(T^a T^b)}{8 \pi^2 x^4} \,,
 }
for some constant $\tau$, then the real mass deformation on $S^3$ is given by \cite{Closset:2012vg}
 \es{RealMassS3}{
   \int d^3 \vec{x}\, \sqrt{g(\vec{x})} \tr \left[ m \left(\frac{i}{r} J(\vec{x}) + K(\vec{x}) \right) \right]  + O(m^2) \,,
 }
where $m = m^a T^a$ is a Lie-algebra valued mass parameter.  Here, `$\tr$' denotes a positive-definite bilinear form on the Lie algebra, which can be thought of as the trace in a conveniently-chosen representation of $\mf{g}$.   For us, we have the flavor symmetry algebra $\mf{g} = \mf{su}(4)$, and we consider a basis of this algebra such that $\tr (T^a T^b) = \delta^{ab}$.   For convenience, we will take `$\tr$' to be the trace in the fundamental (${\bf 4}$) of $\mf{su}(4)$. Note that $\tau$ is related to the stress tensor two-point function coefficient $c_T$ in the $\mathcal{N}=8$ theory by
 \es{tautoCt}{
 \tau=\frac{c_T}{16}\,,
 }
 where our normalization is such that $c_T=1$ for a free real scalar.  For simplicity, let the radius of $S^3$ be set to $r=1$ from now on.

Due to the $\mf{su}(4)$ symmetry, the $S^3$ free energy can be expanded in terms of $\mf{su}(4)$ Casimirs.  For instance,  the first few terms in the expansion at small mass are
 \es{FreeExpansion}{
  F = F_0 + F_2 \tr (m^2) + F_3 \tr (m^3) + \left[F_{4,1} \left( \tr (m^4) - \frac 14  \left(\tr (m^2) \right)^2 \right) + F_{4,2} \left(\tr (m^2) \right)^2 \right] + \cdots \,.
 }
We can of course obtain the same information without the need to consider a completely general $\mf{su}(4)$ mass matrix $m$, and instead focus on a Cartan subalgebra.  Let us order the $T^a$ such that the first three ($a = 1,2 ,3$) correspond to a Cartan subalgebra given explicitly by
 \es{ExplicitTa}{
  T^1 = \diag \left\{\frac 12, \frac 12, -\frac12, -\frac 12\right\} \,, \\
  T^2 = \diag \left\{\frac 12, -\frac 12, \frac12, -\frac 12\right\} \,, \\
  T^3 = \diag \left\{\frac 12, -\frac 12, -\frac12, \frac 12\right\} \,.
 }
With $\tr (T^a T^b) = \delta^{ab}$, the expression \eqref{FreeExpansion} becomes a sum of polynomials in $m_a$, $a = 1,2 ,3$ that are invariant under the action of the Weyl group:
 \es{FreeExpansionAgain}{
  F &= F_0 + F_2 (m_1^2 + m_2^2 + m_3^2)
   + 3 F_3 m_1 m_2 m_3 \\
   &{}+ \bigl[ F_{4, 1} (m_1^2 m_2^2 + m_1^2 m_3^2 + m_2^2 m_3^2)
    + F_{4, 2} (m_1^2 + m_2^2 + m_3^2)^2 \bigr] + \cdots \,.
 }

From \eqref{RealMassS3}, we see that $n$ derivatives of $F(m_a)$ computes integrated $n$-point functions of $J(\vec x)$ and $K(\vec x)$ on $S^3$, possibly supplemented by integrated lower-point functions coming from the $O(m^2)$ terms in \eqref{RealMassS3}. For $n=2$, the second derivative of $F$ gives only the integrated two-point function of $i J + K$ because in this case the $O(m^2)$ terms omitted from \eqref{RealMassS3} could only contribute an integrated one-point function, which vanishes in any CFT\@.  From \eqref{NormTwoPoint}, transformed to $S^3$, the two-derivative of $F$ is evaluated to\footnote{We have $\frac{\partial^2F}{\partial m_a^2} = \int d^3\vec{x}\, d^3 \vec{y}\, \sqrt{g(x)} \sqrt{g(y)} \left[ \langle J^a(\vec{x}) J^a(\vec{y}) \rangle -  \langle K^a(\vec{x}) K^a(\vec{y}) \rangle \right]$.  We can plug in \eqref{NormTwoPoint} converted to $S^3$ and use $\int d^3\vec{x}\, d^3 \vec{y}\, \sqrt{g(x)} \sqrt{g(y)} \frac{\Omega(\vec{x})^{-\Delta} \Omega(\vec{y})^{-\Delta} }{\abs{\vec{x}  - \vec{y}}^{2 \Delta}} = \frac{4^{2- \Delta} \pi^{7/2} \Gamma(\frac 32 - \Delta)}{\Gamma(3 - \Delta)}$ \cite{Cardy:1988cwa}.} \cite{Closset:2012vg}
 \es{cTfromF}{
\frac{\partial^2F}{\partial m_a^2}=\frac{\pi^2}{2}\tau=\frac{\pi^2}{32}c_T\,,
 }
for any $a=1,2,3$.  We would like to perform a similar calculation in the $n=4$ case, first ignoring the $O(m^2)$ contributions in \eqref{RealMassS3}.  (We will return to these contributions in Section~\ref{MSQUARED}.)  Because our final goal is to determine the integrated four-point functions of the $\mathcal{N}=8$ operators $S_{IJ}$ and $P_{AB}$, we must first relate $S_{IJ}$ and $P_{AB}$  to the $\mathcal{N}=2$ operators $J^a$ and $K^a$.
 
The operators $J^a$, $K^a$, $j_\mu^a$ all arise from the stress tensor multiplet of the $\cN = 8$ SCFT, in particular from certain components of the ${\bf 35}_c$ scalars, ${\bf 35}_s$ pseudo-scalars, and of the $\mf{so}(8)$ $R$-symmetry current, respectively.  Indeed, under the subalgebra $\mf{su}(4) \oplus \mf{u}(1)_R \subset \mf{so}(8)_R$ defined by the decompositions of the fundamental representations\footnote{In our convention, the supercharges of the $\cN = 8$ theory transform in ${\bf 8}_v$.  This fixes our embedding of $\mf{su}(4) \oplus \mf{u}(1) \subset \mf{so}(8)$ if we wish to preserve only $\cN = 2$ supersymmetry. The ${\bf 8}_c$ and ${\bf 8}_s$ must then decompose as indicated, or the decompositions may be flipped.  We choose the convention in which the decompositions are as in \eqref{DecompBasic}.}
 \es{DecompBasic}{
  {\bf 8}_v &\to {\bf 6}_0 \oplus {\bf 1}_1 \oplus {\bf 1}_{-1} \,, \\ 
  {\bf 8}_c &\to {\bf 4}_{\frac12} \oplus \overline{\bf 4}_{-\frac12} \,, \\ 
  {\bf 8}_s &\to {\bf 4}_{-\frac12} \oplus \overline{\bf 4}_{\frac12} \,, 
 }
the components of the stress tensor multiplet decompose as:
 \es{StressDecomp}{
  S_{IJ}: \qquad &{\bf 35}_c \to {\bf 10}_1 \oplus \overline{\bf 10}_{-1} \oplus {\bf 15}_0 \,, \\
  \chi^\alpha_{AI}: \qquad &{\bf 56}_v \to {\bf 10}_0 \oplus \overline{\bf 10}_0 \oplus {\bf 15}_1 \oplus {\bf 15}_{-1} \oplus {\bf 6}_0 \,, \\
  P_{AB}: \qquad &{\bf 35}_s \to {\bf 10}_{-1} \oplus \overline{\bf 10}_{1} \oplus {\bf 15}_0 \,, \\
  j^\mu_{ab}: \qquad &{\bf 28} \to {\bf 15}_0 \oplus {\bf 6}_1 \oplus {\bf 6}_{-1} \oplus {\bf 1}_{0}\,, \\
  \psi_a^{\mu \alpha}: \qquad &{\bf 8}_v \to {\bf 6}_0 \oplus {\bf 1}_1 \oplus {\bf 1}_{-1} \,, \\
  T^{\mu\nu}: \qquad &{\bf 1} \to {\bf 1}_0 \,,
 }
with operators in the same $\mf{su}(4)$ representation belonging to the same $\cN = 2$ superconformal multiplet.  Therefore, the $\cN = 2$ flavor current multiplet consists of those components of $S_{IJ}$, $\chi^\alpha_{AI}$, $P_{AB}$, and $j^\mu_{ab}$ that transform in the ${\bf 15}$ of $\mf{su}(4)$ in the decomposition \eqref{StressDecomp}.

We can be more concrete.  Given a generator $T$ of $\mf{su}(4)$ presented as a $4\times 4$ hermitian traceless matrix as, for instance, the generators in \eqref{ExplicitTa}, we can ask which linear combination of the $S_{IJ}$ and which linear combination of the $P_{AB}$ correspond to it.  Let us first focus on $S_{IJ}$.  First, notice that the decomposition ${\bf 8}_c \to  {\bf 4}_{\frac12} \oplus \overline{\bf 4}_{-\frac12} $ implies that if in the fundamental representation of $\mf{su}(4)$,
 \es{TExpansion}{
  T = i (A + i B)
 }
then the same generator acting in the ${\bf 8}_c$ irrep of $\mf{so}(8)$ can be taken to be equal to
 \es{TExpansion2}{
  \tilde T = i \left(A \otimes {\bf 1} + B \otimes \veps \right) \,,
 }
where $\veps \equiv i \sigma_2$.\footnote{We can check this claim as follows.  Since $T$ is hermitian, then $A$ is an anti-symmetric real matrix and $B$ is a symmetric real matrix, which implies that the generator $\tilde T$ is also hermitian.  Then, if $[T^a, T^b] = i f^{abc} T^c$, we can immediately infer that
 \es{ABRelations}{
  [A^a, A^b] - [B^a, B^b] = f^{abc} A^c \,, \qquad
   [A^a, B^b] + [B^a, A^b] = f^{abc} B^c \,,
 } 
which can be used to check that the $\tilde T$ obey the right commutation relations:
 \es{Ttilde}{
  [\tilde T^a, \tilde T^b] = - \left( [A^a, A^b] - [B^a, B^b] \right) \otimes {\bf 1} - \left(   [A^a, B^b] + [B^a, A^b] \right) \otimes \veps
   = i f^{abc} \tilde T^c \,.
 } 
 }
Using the same ingredients, one can also construct a {\em symmetric traceless} matrix $T_{{\bf 35}_c}$ representing the generator $T$ inside the ${\bf 35}_c$:
 \es{SymTraceless}{
  T_{{\bf 35}_c} = A \otimes \veps - B \otimes {\bf 1} \,,
 }
which satisfies $[\tilde T^a, T_{{\bf 35}_c}^b] = i f^{abc} T_{{\bf 35}_c}^c$, so that $T_{{\bf 35}_c}$ indeed corresponds to states in the ${\bf 15}$ of $\mf{su}(4)$.  For the Cartan elements in \eqref{ExplicitTa}, we have
 \es{Cartan35c}{
  T_{{\bf 35}_c}^1 = \frac 12 \diag \left\{1, 1, 1, 1, -1, -1, -1, -1 \right\} \,, \\
  T_{{\bf 35}_c}^2 = \frac 12 \diag \left\{1, 1, -1, -1, 1, 1, -1, -1 \right\} \,, \\
  T_{{\bf 35}_c}^3 = \frac 12 \diag \left\{1, 1, -1, -1, -1, -1, 1, 1 \right\} \,.
 }
This implies that $J^a \propto (T_{{\bf 35}_c}^a)^{IJ} S_{IJ}$ are given by
 \es{JDef}{
  J^1 &= N_J \left[ S_{11} + S_{22} + S_{33} + S_{44} - S_{55} - S_{66} - S_{77} - S_{88} \right] \,, \\
  J^2 &= N_J \left[ S_{11} + S_{22} - S_{33} - S_{44} + S_{55} + S_{66} - S_{77} - S_{88} \right] \,, \\
  J^3 &= N_J \left[ S_{11} + S_{22} - S_{33} - S_{44} - S_{55} - S_{66} + S_{77} + S_{88} \right] \,, 
 }
where the normalization constant $N_J$ is determined to be 
 \es{GotNJ}{
  N_J = \sqrt{\frac{\tau}{128 \pi^2}} = \sqrt{\frac{c_T}{2^{11} \pi^2}} 
 }
such that the normalization \eqref{NormTwoPoint} is obeyed.

A similar procedure can be repeated to give $K^a$, but we have to be careful that the $\mf{su}(4)$ generators written as $8 \times 8$ matrices in the ${\bf 8}_s$ irrep are consistent with the symbols $E_{aIA}$ defined in the previous section.  One can check that this is indeed the case for our choice of $E_{aIA}$, and that the $\mf{su}(4)$ generators $T$ in \eqref{TExpansion} are also represented by the $\tilde T$ in \eqref{TExpansion2} in the ${\bf 8}_s$ representation.\footnote{What needs to be checked is that $\tilde T^{IJ} = 4 E^{aI}{}_A E_{a}{}^J{}_B \tilde T^{AB}$, where $\tilde T^{IJ}$ and $\tilde T^{AB}$ are components of the matrices \eqref{TExpansion2}.}  Then, by analogy with \eqref{JDef}, the three $K^a$ corresponding to \eqref{ExplicitTa} are
 \es{KDef}{
  K^1 &= \sqrt{2}N_J \left[ P_{11} + P_{22} + P_{33} + P_{44} - P_{55} - P_{66} - P_{77} - P_{88} \right] \,, \\
  K^2 &=  \sqrt{2}N_J \left[ P_{11} + P_{22} - P_{33} - P_{44} + P_{55} + P_{66} - P_{77} - P_{88} \right] \,, \\
  K^3 &=  \sqrt{2}N_J \left[ P_{11}  + P_{22} - P_{33} - P_{44} - P_{55} - P_{66} + P_{77}  + P_{88} \right] \,,
 }
where the normalization was chosen such that \eqref{NormTwoPoint} is obeyed.

Having determined the relations between $(J^a, K^a)$ and $(S_{IJ}, P_{AB})$ in Eqs.~\eqref{JDef} and \eqref{KDef}, as well as the way $(J^a, K^a)$ appear in the mass deformed theory (Eq.~\eqref{RealMassS3}), we can then relate the various mass derivatives of the $S^3$ free energy to integrated 4-point correlators of $S_{IJ}$ and $P_{AB}$.  The non-zero derivatives are of the form
 \es{d4Fdm4}{
  \frac{\partial^4 F}{\partial m_a^2 \partial m_b^2} &=  -N_J^4 
   \Bigl(  s_{ab}^k I^3_{1,1}\left[{\cal S}_k^\text{conn}(U, V)\right] 
    + p_{ab}^k I^3_{2,2}\left[{\cal P}_k^\text{conn}(U, V)\right]  +  r_{ab}^k I^3_{1,2}\left[{\cal R}_k^\text{conn}(U, V)\right] \Bigr)
 }
where $a, b = 1, 2, 3$, for now on we set $r=1$, and we define the integrated quantity
 \es{4pointS3Int}{
I^d_{\Delta_A, \Delta_B}[{\cal G}] = \int  \left( \prod_{i=1}^4 d^d\vec{x}_i \right)  \frac{ \left[  \Omega(\vec{x}_1) \Omega(\vec{x}_2) \right]^{d-\Delta_A}  \left[  \Omega(\vec{x}_3) \Omega(\vec{x}_4) \right]^{d-\Delta_B} }{x_{12}^{2\Delta_A}x_{34}^{2\Delta_B}}\mathcal{G}(U,V)\,,
 }  
which computes a four point function $\langle AABB\rangle$ integrated over $S^d$ for any dimension $d$. If $a=b$, then the coefficients $s_{ab}^k$, $p_{ab}^k$, $r_{ab}^k$ are
 \es{sprFirst}{
  s_{aa}^k =  \frac{p_{aa}^k}{4}  = \begin{pmatrix} 64 & 64 & 64 & 8 & 8 & 8 \end{pmatrix}  \,, \qquad
   r_{aa}^k = \begin{pmatrix} -768 & -84 & 0 \end{pmatrix} \,,
 }
while if $a \neq b$, then
 \es{sprSecond}{
  s_{ab}^k = \frac{p_{ab}^k}{4} = \begin{pmatrix} 64 & 0 & 0 & 8 & 8 & 8 \end{pmatrix} \,, \qquad
   r_{ab}^k = \begin{pmatrix} 
    - 256 & 4 & 0 
   \end{pmatrix}  \,.
 }

 \subsection{Integrated four point function}
 \label{S3}
 
Let us now evaluate more explicitly the quantity $I^d_{\Delta_A,\Delta_B}$ defined in \eqref{4pointS3Int}.  While we are mainly interested in the case $d = 3$ and $\Delta_{A, B} = 1$ or $2$, we nevertheless keep $d$, $\Delta_A$, and $\Delta_B$ completely general in this section.  Eq.~\eqref{4pointS3Int} contains $4d$ integrals as written, but using conformal symmetry one can perform $4d-2$\footnote{For $d=2,3$ these correspond to the $\frac{(d+1)(d+2)}{2}$ generators of the conformal group $SO(d,2)$, while for $d\geq4$, after we have used conformal transformation to set one point to the origin and the other to infinity, there is a nontrivial stability group $SO(d-2)$ so that the number of integrals we can perform is $\frac{(d+1)(d+2)}{2}-\frac{(d-3)(d-2)}{2}=4d-2$.} of them, as follows.   The first step is to notice that the integral \eqref{4pointS3Int} is rotationally-invariant on $S^d$, so one can rotate the point $\vec{x}_4$ to any fixed point of our choosing $\vec{x}_4 = \vec{x}_{4*}$, using
 \es{x4Formula}{
  \int d^d \vec{x}_4\,  \Omega(\vec{x}_4)^d \, f\left( \left[\Omega(\vec{x}_4) \Omega(\vec{x}_i)\right]^{\frac 12}(\vec{x}_4 - \vec{x}_i) \right) = \Vol(S^d) f\left( \left[\Omega(\vec{x}_{4*}) \Omega(\vec{x}_i)\right]^{\frac 12}(\vec{x}_{4*} - \vec{x}_i) \right) \,,
 }
 where $\Vol(S^d)=\frac{2\pi^{\frac{d+1}{2}}}{\Gamma\left[\frac{d+1}{2}\right]}$.
A convenient choice is $\abs{\vec{x}_{4*}} = \infty$, which gives
   \es{4pointS3Int1}{
I^d_{\Delta_A, \Delta_B}[{\cal G}] = \frac{\Vol(S^d)}{4^{\Delta_B}}\int \left( \prod_{i=1}^3 d^d \vec{x}_i \right)    \frac{\left[  \Omega(\vec{x}_1) \Omega(\vec{x}_2) \right]^{d-\Delta_A}  \left[  \Omega(\vec{x}_3) \right]^{d-\Delta_B} }{x_{12}^{2\Delta_A}}\mathcal{G}\left(\frac{x_{12}^2}{x_{13}^2},\frac{x_{23}^2}{x_{13}^2}\right)\,.
 }  
We can then translate $\vec{x}_1\to \vec{x}_1+\vec{x}_3$ and $\vec{x}_2\to \vec{x}_2+\vec{x}_3$, and write \eqref{4pointS3Int1} as
\es{4pointS3Int2}{
I^d_{\Delta_A, \Delta_B}[{\cal G}] = \frac{\Vol(S^d)}{4^{\Delta_B}}\int  \left( \prod_{i=1}^3 d^d \vec{x}_i \right)        \frac{\left[  \Omega(\vec{x}_1 + \vec{x}_3) \Omega(\vec{x}_2 + \vec{x}_3) \right]^{d-\Delta_A}  \left[  \Omega(\vec{x}_3) \right]^{d-\Delta_B} }{x_{12}^{2\Delta_A}}\mathcal{G}\left(\frac{x_{12}^2}{x_{1}^2},\frac{x_{2}^2}{x_{1}^2}\right)\,.
 }  
 Note that the $\vec{x}_3$ dependence is only in the prefactor now.  We can then use the remaining rotational symmetry to set $\vec{x}_1$ and $\vec{x}_2$ to $\vec{x}_{1*} = (r_1, 0, \dots,0)$ and $\vec{x}_{2*} = (r_2 \cos \theta, r_2 \sin \theta, 0,\dots,0)$, respectively:
 \es{4pointS3Int3}{
&I^d_{\Delta_A, \Delta_B}[{\cal G}] = \frac{\Vol(S^d)\Vol(S^{d-1})\Vol(S^{d-2})}{4^{\Delta_B}}\int  d^d \vec{x}_3 \, dr_1 dr_2\, d\theta \, r_1^{d-1} r_2^{d-1} \sin^{d-2} \theta\\
&\times \frac{\left[   \Omega(\vec{x}_{1*} + \vec{x}_3) \Omega(\vec{x}_{2*} + \vec{x}_3) \right]^{d-\Delta_A}  \left[  \Omega(\vec{x}_3) \right]^{d-\Delta_B}  }{\abs{\vec{x}_{1*} - \vec{x}_{2*} }^{2\Delta_A}} \mathcal{G}\left(\frac{r_1^2 + r_2^2 - 2 r_1 r_2 \cos \theta}{r_1^2},\frac{r_2^2}{r_1^2}\right)\,.
 }  
 
Then we can change variables from $(r_1, r_2, \vec{x}_3)$ to $(z_0, r, \vec{z})$ defined through $r_1 = 2/z_0$, $r_2 = 2r/z_0$, $\vec{x}_3 = 2 \vec{z}/z_0$, after which the integral takes the form
 \es{IRewrite}{
 & I^d_{\Delta_A, \Delta_B}[{\cal G}] = \frac{\Vol(S^d)\Vol(S^{d-1})\Vol(S^{d-2})}{2^{2\Delta_A + 2\Delta_B-3d}} \int dr\, d\theta\, r^{d-1} \sin^{d-2} \theta   \frac{\mathcal{G}\left(1 + r^2 - 2 r \cos \theta,r^2\right)}{(1 + r^2 - 2 r \cos \theta)^{\Delta_A}} \\
  &\times\lim_{\vec x_4'\to\infty}\left[\abs{\vec{x}_4'}^{2(d - \Delta_B)} \int \frac{d^d\vec z dz_0}{z_0^{d+1}}  G^{d-\Delta_A}_{B\partial}(z,\vec x_1') G^{d-\Delta_A}_{B\partial}(z,\vec x_2') G^{d-\Delta_B}_{B\partial}(z,\vec x_3') G^{d-\Delta_B}_{B\partial}(z,\vec x_4')\right] \,,\\
&   \qquad\quad\,\,    \vec x_1'=(1, 0,\dots, 0)\,,\qquad     \vec x_2'= (r \cos \theta, r \sin \theta, 0,\dots,0)\,,\qquad     \vec x_3'=(0, \dots, 0)\,, 
 } 
where $z = (z_0, \vec{z})$ and $G_{B\partial}^r(z, \vec{x})$ is the AdS bulk-to-boundary propagator defined in \eqref{Dfunc2}. The quantity in square brackets in \eqref{IRewrite} can be written in terms of the $\bar D$ function described in Appendix \ref{dbarApp}, so that we get
  \es{IRewrite2}{
 & I^d_{\Delta_A, \Delta_B}[{\cal G}] =
    \frac{\Vol(S^d)\Vol(S^{d-1})\Vol(S^{d-2})\pi^{\frac d2}\Gamma( \frac {3d}{2}- \Delta_A - \Delta_B)}{2^{2\Delta_A + 2\Delta_B-3d+1} \Gamma(d - \Delta_A)^2 \Gamma(d - \Delta_B)^2} \\
&\times     \int dr\, d\theta\,  r^{d-1} \sin^{d-2} \theta
      \left[\bar D_{d - \Delta_A, d - \Delta_A, d - \Delta_B, d - \Delta_B}(U, V)\frac{ \mathcal{G}\left(U, V\right)}{U^{\Delta_A}}\right]_{\substack {U=1 + r^2 - 2 r \cos \theta \\ V=r^2}}\,.
  }
Note that this formula is symmetric under interchanging $\Delta_A \leftrightarrow \Delta_B$, because the $\bar D$ functions obey the relation $\bar D_{aabb}(U, V) = U^{b-a} \bar D_{bbaa}(U,V)$ \cite{Arutyunov:2002fh}.  Combined with \eqref{d4Fdm4}, the formula \eqref{IRewrite2} allows for an explicit evaluation of the integrated four-point functions provided that we know the functions of $U$ and $V$ appearing in \eqref{SPFourPoint} and \eqref{SSPP}.  We will determine these functions in the $1/c_T$ expansion in the next section.

\subsection{Order $m^2$ terms}
\label{MSQUARED}

We derived the formula~\eqref{d4Fdm4} under the assumption that there are no $O(m^2)$ terms in \eqref{RealMassS3}.  However, if there are $\Delta = 1$ scalar operators ${\cal O}$ present, then an $m^2 {\cal O}$ term is possible in \eqref{RealMassS3} (and generally present).  Such a term would contribute additively to \eqref{d4Fdm4}.

\subsubsection{Order $m^2$ terms in the free theory}

The first example where there are $O(m^2)$ terms accompanying the mass deformation \eqref{RealMassS3} is a free ${\cal N} = 8$ theory.  Such a theory has a presentation in terms in terms of eight real scalars fields $X_I$ and eight Majorana fermions $\psi_A$.  Equivalently, we can group the eight real scalars into four complex combinations $Z_i = \frac{X_{2i-1} + i X_{2i}}{\sqrt{2}}$ and similarly for the fermions, so that the free theory action is 
 \es{FreeAction}{
  S_\text{free} = \int d^3x \, \partial_\mu Z_i \partial^\mu \bar Z^i + \text{(fermions)}\,.
 }
In this case we have
 \es{JFree}{
  J^1 &= \frac 12 \left( \abs{Z_1}^2 + \abs{Z_2}^2 - \abs{Z_3}^2 - \abs{Z_4}^2 \right) \,, \\
  J^2 &=\frac 12 \left( \abs{Z_1}^2 - \abs{Z_2}^2 + \abs{Z_3}^2 - \abs{Z_4}^2 \right) \,, \\
  J^3 &= \frac 12 \left(\abs{Z_1}^2 - \abs{Z_2}^2 - \abs{Z_3}^2 + \abs{Z_4}^2 \right) \,,
 }
and indeed from $\langle Z_i(\vec{x}) \bar Z^j(0) \rangle = \frac{\delta_i^j}{4 \pi \abs{\vec{x}}}$ we derive  $\langle J^a(\vec{x}) J^b(0) \rangle = \frac{\delta^{ab}}{16 \pi^2 x^2}$, in agreement with \eqref{NormTwoPoint}, \eqref{tautoCt}, and $c_T = 16$.  The mass deformation \eqref{RealMassS3} implies that the scalars $Z_i$ have masses $M_i$ given by
 \es{GotMi}{
  M_i = \left(\frac{m_1 + m_2 + m_3}{2}, \frac{m_1 - m_2 - m_3}{2}, \frac{-m_1 + m_2 - m_3}{2}, \frac{-m_1 - m_2 + m_3}{2} \right) \,.
 }   
The real mass deformation on $S^3$ has both a linear and a quadratic term in $M_i$, namely
 \es{SMassFree}{
  S_\text{mass} = \int d^3x\, \sqrt{g} \sum_{i=1}^4 \left(i  M_i \abs{Z_i}^2 + M_i^2 \abs{Z_i}^2 \right) + \text{(fermions)} \,,
 }
and so we have\footnote{This expression can also be written as 
$$
 S_\text{mass} = \int d^3 x\, \sqrt{g}  \tr \left[  i  m J  + m^2 J + m^2 {\cal O}_S \right] + \text{(fermions)}
$$
in a manifestly $\mf{su}(4)$-invariant form.} \cite{Freedman:2013oja}
 \es{SMassFreeAgain}{
 S_\text{mass} &= \int d^3x\, \sqrt{g}  \Biggl[ \left(  im_1 + m_2 m_3 \right) J^1
  + \left(  im_2 + m_1 m_3\right) J^2 + \left(  im_3 + m_1 m_2 \right) J^3  \\
  {}&+  (m_1^2 + m_2^2 + m_3^2) {\cal O}_S \Biggr]  + \text{(fermions)} \,,
 }
where
 \es{OSDef}{
  {\cal O}_S \equiv \frac{1}{4} \left( \abs{Z_1}^2 + \abs{Z_2}^2 + \abs{Z_3}^2 + \abs{Z_4}^2 \right)
 } 
is an  $\mf{su}(4)$ singlet.  Thus
 \es{dFFree}{
  \frac{\partial^4 F}{\partial m_a^2 \partial m_b^2} = \text{\eqref{d4Fdm4}}\ {}- 
  \begin{cases}
   12 \int \left( \langle \cO_S \cO_S \rangle
    + \langle \cO_S J^a J^a \rangle \right) & \text{if $a=b$} \,, \\
   4 \int \left( \langle \cO_S \cO_S \rangle + \langle \cO_S J^a J^a \rangle 
    + \frac12 \langle J^c J^c \rangle + \langle J^a J^b J^c \rangle \right) & \text{if $a \neq b$} \,,
   \end{cases}
 }
where $c \neq a$ and $c \neq b$. We can then use \cite{Cardy:1988cwa}
 \es{IntegratedCorrelators}{
  \int d^3 \vec{x} \, d^3 \vec{y} \, \sqrt{g(\vec{x})} \sqrt{g(\vec{y})} \frac{\Omega^{-1}(\vec{x}) \Omega^{-1}(\vec{y})}{\abs{\vec{x} - \vec{y}}^2} &= 4 \pi^4 \,, \\
   \int d^3 \vec{x} \, d^3 \vec{y} \,d^3 \vec{z} \, \sqrt{g(\vec{x})} \sqrt{g(\vec{y})} \sqrt{g(\vec{z})} \frac{\Omega^{-1}(\vec{x}) \Omega^{-1}(\vec{y}) \Omega^{-1}(\vec{z})}{\abs{\vec{x} - \vec{y}}
    \abs{\vec{x} - \vec{z}} \abs{\vec{y} - \vec{z}}} &= 16 \pi^5 \,,
 }
as well as the correlators
 \es{Correlators}{
  \langle J^c(\vec{x})  J^c(\vec{y}) \rangle &= 4\langle {\cal O}_S(\vec{x})  {\cal O}_S(\vec{y}) \rangle   =  \frac{1}{(4 \pi)^2}\frac{\Omega^{-1}(\vec{x}) \Omega^{-1}(\vec{y})}{\abs{\vec{x} - \vec{y}}^2} \,, \\
  \langle J^a (\vec{x})  J^b(\vec{y}) J^c(\vec{z}) \rangle &= 2 \langle {\cal O}_S(\vec{x})  J^a(\vec{y}) J^a(\vec{z}) \rangle 
   = \frac{1}{(4 \pi)^3} \frac{\Omega^{-1}(\vec{x}) \Omega^{-1}(\vec{y}) \Omega^{-1}(\vec{z})}{\abs{\vec{x} - \vec{y}}
    \abs{\vec{x} - \vec{z}} \abs{\vec{y} - \vec{z}}} 
 }
that hold whenever $a$, $b$, $c$ are all distinct, to find 
 \es{dFFreeAgain}{
  \frac{\partial^4 F}{\partial m_a^2 \partial m_b^2} = \text{\eqref{d4Fdm4}}\ -  
  \frac{9 \pi^2}{4} 
 }
regardless of whether $a = b$ or $a \neq b$.  The $O(m^2)$ terms in \eqref{RealMassS3} thus contribute $-\frac{9\pi^2}{4}$ to the fourth mass derivatives \eqref{dFFreeAgain}.

For completeness, let us note that in the free theory, we have ${\cal R}_k^\text{conn} = 0$ and 
 \es{SPFree}{
  {\cal S}_k^\text{conn} &= \begin{pmatrix} 0 & 0 & 0 & \frac{4U}{\sqrt{V}} & 4 \sqrt{\frac{U}{V}} & 4 \sqrt{U} 
  \end{pmatrix} \,, \\
  {\cal P}_k^\text{conn} &= \begin{pmatrix}
   0 & 0 & 0 & \frac{U^2}{V^{3/2}} (U -V - 1) & - \frac{\sqrt{U}}{V^{3/2}}(U + V - 1) & \sqrt{U} (V - U - 1) 
   \end{pmatrix} \,.
  }
The integrals in \eqref{d4Fdm4} can be performed analytically with the result  
 \es{FourPointFree}{
  \text{\eqref{d4Fdm4}}\ = - \frac{\pi^4}{4}  + \frac{9 \pi^2}{4} \,.
 }
Combining this expression with \eqref{dFFreeAgain}, we find that
 \es{derFinalFree}{
   \frac{\partial^4 F}{\partial m_a^2 \partial m_b^2} =- \frac{\pi^4}{4} \,.
 }
This result is indeed correct and serves as a check of our formalism.  Indeed, in a free theory, the mass-deformed $S^3$ free energy can be computed by directly evaluating the required Gaussian integrals.  As an alternative, one can use the supersymmetric localization result of \cite{Jafferis:2010un} that gives
 \es{FFreeLoc}{
  F = -\sum_{i=1}^4 \ell\left(\frac 12 + M_i \right) \,,
 }
with the function $\ell(z)$ defined in (1.3) of \cite{Jafferis:2010un}.  It is straightforward to see that \eqref{FFreeLoc} implies \eqref{derFinalFree} for any $a, b = 1, 2, 3$.

\subsubsection{Order $m^2$ terms at strong coupling}

In a strongly coupled CFT with a holographic dual, as will be the case we study in the next section, one also expects a term of order $m^2$ similar to \eqref{SMassFreeAgain} to supplement \eqref{RealMassS3}.  One difference is that in a generic strongly coupled CFT there is no operator ${\cal O}_S$ of dimension~$1$ so we can drop the contributions involving ${\cal O}_S$ from \eqref{dFFree}.  In addition, the correlators $ \langle J^c(\vec{x})  J^c(\vec{y}) \rangle$ and $\langle J^a (\vec{x})  J^b(\vec{y}) J^c(\vec{z}) \rangle$ are $c_T/16$ times larger than in the free theory, so \eqref{dFFreeAgain} becomes
\es{dFIntAgain}{
  \frac{\partial^4 F}{\partial m_a^2 \partial m_b^2} = \text{\eqref{d4Fdm4}}\ -  
   \begin{cases} 
    0 & \text{if $a=b$} \,, \\
    \frac{3 \pi^2}{64} c_T  & \text{if $a \neq b$} \,.
    \end{cases}
 }
One notable feature of this result is that while both $\frac{\partial^4 F}{\partial m_a^2 \partial m_b^2}$ and \eqref{d4Fdm4} have non-trivial expansions in $1/c_T$, the last term of \eqref{dFIntAgain} is simply proportional to $c_T$.  For this reason, it will not play any role in the next section.

\section{From ABJM to the 11d S-matrix}
\label{ABJM}

We will now apply the machinery of the previous section to fix the coefficients $B^4_i$ and $B^6_i$ of the $c_T^{-\frac53}$ and $c_T^{-\frac{19}{9}}$ terms in the large $c_T$ expansion of the Mellin amplitude $M^{SSSS}$ \eqref{MellinIntro} (Mellin amplitudes are reviewed in Appendix \ref{MELLIN}), which in the flat space limit yield the $R^4$ and $D^4R^4$ terms, respectively, in the 11d S-matrix.  Given a correlator $\langle {\cal O}_1(\vec{x}_1) \cO_2(\vec{x}_2) \cO_3(\vec{x}_3) \cO_4(\vec{x}_4) \rangle$ of operators in the stress tensor multiplet of ABJM theory with the corresponding Mellin amplitude $M^{\cO_1 \cO_2 \cO_3 \cO_4}(s, t)$, then using \cite{Fitzpatrick:2011hu} we can deduce that the flat space four supergraviton scattering amplitude with momenta restricted to 4 dimensions is (up to an overall $\Delta_i$-independent numerical factor) \cite{Chester:2018aca}:
 \es{FlatFromCFT}{
  {\cal A}(\eta_i, s, t) &=  \Gamma\left(\frac12\sum_i\Delta_i-\frac32 \right) \left[ \int_{S^7} d^7x \, \sqrt{g} \prod_{i=1}^4 \Psi_{\eta_i}^{\cO_i} (\vec{n})  \right] \\
  &{}\times
  \lim_{L \to \infty} (2L)^7 \int_{c-i\infty}^{c+ i \infty} d\alpha\, e^\alpha \alpha^{\frac32-\frac12\sum_i\Delta_i} M^{\cO_1 \cO_2 \cO_3 \cO_4}\left(\frac{L^2}{2 \alpha} s,  \frac{L^2}{2 \alpha} t \right)\,.
 }
Here $\eta_i$ are the polarizations of the supergravitons, and the factor in the square brackets is a form factor involving the wavefunctions $\Psi_{\eta_i}^{\cO_i}(\vec{n})$ of the modes dual to the operators ${\cal O}_i$ in the internal unit $S^7$.  The integration contour in \eqref{FlatFromCFT} must have $c>0$.  From \eqref{FlatFromCFT} as well as the leading order AdS/CFT relation
 \es{cTAdS}{
  c_T = \frac{64}{3\pi} \sqrt{2k} N^{3/2} + \cdots = \frac{2^{11}}{3 \pi k} \frac{L^9}{\ell_p^9}+ \cdots\,,
 }
it can be seen that at order $1/c_T^n$ it is only the terms that at large $s$ and $t$ scale as $s^a t^b$ with $a+b={(9n-7)/2}$ that contribute to \eqref{FlatFromCFT}. (The AdS radius $L$ drops out for terms with this growth;  terms with slower growth come suppressed in $L$ and drop out after taking $L \to \infty$.)  They contribute a homogeneous term of order $(9n-7)/2$ in the Mandelstam invariants multiplied by $\ell_p^{9n-7}$.  Because the leading supergravity term is proportional to $c_T^{-1}$ (and it thus gives a linear scattering amplitude), the $R^4$ contribution to $M$ will scale as $c_T^{-\frac 53}$ (and it gives a quartic scattering amplitude), while the potential $D^4 R^4$ contribution that we want to show is absent scales as $c_T^{-\frac{19}{9}}$ (and would give a sextic scattering amplitude).

We will now derive the $c_T^{-1}$, $c_T^{-\frac53}$, and $c_T^{-\frac{19}{9}}$ Mellin amplitudes for $M^{PPPP}$ and $M^{SSPP}$ from $M^{SSSS}$. For $c_T^{-\frac53}$, and $c_T^{-\frac{19}{9}}$, we then convert these expressions to position space and integrate them.  Finally, we can compare these integrated 4-point functions to the appropriate derivatives of $F(m_1,m_2)$ in ABJM theory, which can been computed to all order in $c_T$, to recover the known values of $B^4_i$ and show the new result $B^6_i=0$.

\subsection{Mellin amplitudes for $\langle SSSS\rangle$, $\langle PPPP\rangle$, and $\langle SSPP\rangle$}
\label{moreMellin}

As reviewed in the Introduction, superconformal symmetry fixes the $c_T^{-1}$, $c_T^{-\frac53}$, and $c_T^{-\frac{19}{9}}$ terms $M^{SSSS}_{(1)}$, $M^{SSSS}_{(4)}$, and $M^{SSSS}_{(6)}$, respectively, in $M^{SSSS}$ to take the following form:
\es{MellinOurs}{
M^{SSSS}_{(1)}(s,t)=&B_1^1M_\cS^1\,,\qquad M^{SSSS}_{(4)}(s,t)=B_4^4M_\cS^4+B_1^4M_\cS^1\,,\\
 M^{SSSS}_{(6)}(s,t)=&B_6^6M_\cS^6+B_4^6M_\cS^4+B_1^6M_\cS^1\,,
}
where $M_\cS^p$ are asymptotically degree $p$ polynomials in $s,t$ that we can write in the basis \eqref{SPFourPoint} as
\es{SbasisM}{
M_\cS^{p}(s,t)=& Y_{12}^2 Y_{34}^2 \tilde \cS_{1}^p
   + Y_{13}^2 Y_{24}^2 \tilde \cS_{2}^p +Y_{14}^2 Y_{23}^2 \tilde \cS_{3}^p  \\
    &+ Y_{13} Y_{14} Y_{23} Y_{24} \tilde\cS_{4}^p+ Y_{12} Y_{14} Y_{23} Y_{34} \tilde  \cS_{5}^p
     +Y_{12} Y_{13} Y_{24} Y_{34}  \tilde\cS_{6}^p  \,,
}
where $\tilde\cS_i^p$ are related under crossing as
\es{crossS1}{
\tilde\cS^p_2(s,t)&=\cS^p_1(4-s-t,t)\,,\qquad \tilde\cS^p_3(s,t)=\cS^p_1(t,s)\,,\\ \tilde\cS^p_5(s,t)&=\cS^p_4(4-s-t,t)\,,\qquad \tilde\cS^p_6(s,t)=\cS^p_4(t,s)\,.
}

For $p=1$, \cite{Zhou:2017zaw} computed $M_\cS^1$ as an infinite sum of poles, whose explicit formula we relegate to Appendix \ref{R4D4R4}\@. In the large $s,t$ limit this amplitude is normalized as
   \es{M1}{
M_{\cS,\text{asymp}}^{1} \equiv  M_\cS^{1} \underset{s,t\to\infty}\longrightarrow \frac{1}{stu} \left(tuY_{12}Y_{34}+stY_{13}Y_{24}+suY_{14}Y_{23}\right)^2\,,
   }
   where recall that $u=-s-t$ in the large $s,t$ limit. The asymptotic form $M_{\cS,\text{asymp}}^{1}$ of this AdS$_4$ Mellin amplitude was derived in \cite{Chester:2018aca} from the tree amplitude in 4D ungauged $\mathcal{N}=8$ supergravity \cite{Cremmer:1979up,Cremmer:1978ds,deWit:1977fk}. $M_\cS^1$ includes the contribution of the Mellin space conformal blocks for the stress tensor multiplet, which must be proportional to the OPE coefficient $\lambda^2_\text{Stress}$. On general grounds $\lambda^2_\text{Stress}\propto c_T^{-1}$, so in our normalization we have \cite{Zhou:2017zaw,Chester:2018aca}
   \es{B1}{
   B_1^1=\frac{32}{\pi^2c_T}\,,\qquad B_1^p=0\quad\text{for}\quad p>1\,.
   }

For $p>1$, $M_\cS^p$ are maximum degree $p$ polynomials in $s,t,u$. These terms were computed explicitly in \cite{Chester:2018aca} for $p\leq10$. For instance, for $p=4$ we have
\es{R4Mellin}{
\tilde\cS^4_1&=s^2 t^2-\frac{34 s^2 t}{7}+\frac{40 s^2}{7}+2 s t^3-\frac{90 s t^2}{7}+\frac{932 s t}{35}-\frac{624
   s}{35}+t^4-8 t^3+\frac{116 t^2}{5}-\frac{144 t}{5}+\frac{64}{5}\,,\\
\tilde\cS^4_4&=-2 s^3 t-\frac{36 s^3}{7}-2 s^2 t^2+\frac{124 s^2 t}{7}+\frac{1152 s^2}{35}+\frac{68 s
   t^2}{7}-\frac{352 s t}{7}-\frac{464 s}{7}-\frac{80 t^2}{7}+\frac{320 t}{7}+\frac{1472}{35}\,,
}
where the other $\tilde\cS^4_i$ are given by crossing \eqref{crossS1}. For $M_\cS^{6}$, we relegate the explicit terms to Appendix \ref{R4D4R4}\@. In the large $s,t$ limit these amplitudes are normalized as
   \es{M46}{
  M_\cS^{4} \underset{s,t\to\infty}\longrightarrow stu M_{\cS,\text{asymp}}^{1}\,,\qquad M_\cS^{6} \underset{s,t\to\infty}\longrightarrow stu(s^2+t^2+u^2)M_{\cS,\text{asymp}}^{1}\,,
   }
   where $M_{\cS,\text{asymp}}^{1}$ was given in \eqref{M1}. Note that $stu$ and $stu(s^2+t^2+u^2)$ are the only crossing symmetric polynomials in $s,t,u$ of their respective degree.

We can also write the Mellin amplitudes for $\langle PPPP\rangle$ and $\langle SSPP\rangle$ in the large $c_T$ expansion as \eqref{MellinOurs}, where the corresponding polynomial amplitudes $M_\cP^p$ and $M_\cR^p$ can be written in the same basis as \eqref{SPFourPoint} and \eqref{SSPP}:
\es{PbasisM}{
M_\cP^{p}(s,t)=& X_{12}^2 X_{34}^2 \tilde \cP_{1}^p
   +X_{13}^2 X_{24}^2 \tilde \cP_{2}^p  + X_{14}^2 X_{23}^2 \tilde \cP_{3}^p\\
    &+ X_{13} X_{14} Y_{23} Y_{24}\tilde\cP_{4}^p + X_{12} X_{14} X_{23} X_{34} \tilde  \cP_{5}^p
     + X_{12} X_{13} Y_{24} Y_{34} \tilde\cP_{6}^p  \,,\\
     M_\cR^{p}(s,t)=& Y_{12}^2 X_{34}^2 \tilde \cR_{1}^p
   + \left[ (Y_1 \circ Y_2) \cdot (X_3 \circ X_4) \right]^2 \tilde \cR_{2}^p +  (Y_1 \circ Y_2) \cdot (X_3 \circ X_4) Y_{12} X_{34}\tilde \cR_{3}^p\,,
}
where $\tilde\cP_i^p$ are related under crossing as
\es{crossP1}{
\tilde\cP^p_2(s,t)&=\cP^p_1(8-s-t,t)\,,\qquad \tilde\cP^p_3(s,t)=\cP^p_1(t,s)\,,\\ \tilde\cP^p_5(s,t)&=\cP^p_4(8-s-t,t)\,,\qquad \tilde\cP^p_6(s,t)=\cP^p_4(t,s)\,.
}

In Section~\ref{4points}, we showed in position space that the components $\mathcal{P}_i$ and $\mathcal{R}_i$ of $\langle PPPP\rangle$ and $\langle SSPP\rangle$ are related to $\mathcal{S}_i$ according to the differential operators $\mathcal{D}^R_i(U,V,\partial_U,\partial_V)$ and $\mathcal{D}^P_{ij}(U,V,\partial_U,\partial_V)$ \eqref{wards}, which are rational functions of $U,V$ and degree $2$ and $4$ in $\partial_U,\partial_V$, respectively. From the definition \eqref{mellinDef} we can relate their Mellin amplitudes as
   \es{mellinRelation}{
  \tilde \cR_i(s,t)=&\frac{\Gamma\left[1-\frac s2\right]^2\Gamma\left[1-\frac t2\right]^2\Gamma\left[\frac {s+t}{2}-1\right]^2}{\Gamma\left[1-\frac s2\right]\Gamma\left[2-\frac s2\right]\Gamma\left[\frac32-\frac t2\right]^2\Gamma\left[\frac {s+t}{2}-\frac32\right]^2}\widehat {V^{\frac12}\mathcal{D}_i^R}{(U,V,\partial_U,\partial_V)}\tilde \cS_1(s,t)\,,\\
    \tilde \cP_i(s,t)=&\frac{\Gamma\left[1-\frac s2\right]^2\Gamma\left[1-\frac t2\right]^2\Gamma\left[\frac {s+t}{2}-1\right]^2}{\Gamma\left[2-\frac s2\right]^2\Gamma\left[2-\frac t2\right]^2\Gamma\left[\frac {s+t}{2}-2\right]^2}\widehat {V\mathcal{D}^P_{ij}}{(U,V,\partial_U,\partial_V)}\tilde \cS_j(s,t)\,,
   }
   where the hatted operators act on $\tilde \cS_i(s,t)$ as
  \es{3DMellin}{
\widehat{\partial^m_U} \tilde \cS _{i}(s,t)&=\frac{\Gamma\left[\frac s2+1\right]}{\Gamma\left[\frac s2-m+1\right]}  \widehat U^{-m} \tilde \cS _{i}(s,t)\,,\\
\widehat{\partial^m_V} \tilde  \cS_{i}(s,t)&=\frac{\Gamma\left[\frac t2\right]}{\Gamma\left[\frac t2-m\right]}\widehat V^{-m} \tilde \cS_{i}(s,t)\,,\\
\widehat{U^mV^n} \tilde \cS _{i}(s,t)&=\tilde \cS _{i}(s-2m,t-2n)\left(1-\frac{s}{2}\right)_m^2\left(1-\frac{t}{2}\right)_n^2\left(\frac{s+t}{2}-1\right)_{-m-n}^2\,.
}
Using these formulae and the explicit expressions for $\cS^p_i$ for $p=1,4,6$ in \eqref{sugraMellinS}, \eqref{R4Mellin}, and \eqref{D4R4MellinS}, respectively, we can now derive $\cP^p_i$ and $\cR^p_i$. For instance, for $p=4$ we find
\es{quartic2}{
 \tilde \cP^4_1=&99 s^2 t^2-594 s^2 t+904 s^2+198 s t^3-2178 s
   t^2+7812 s t-\frac{64720 s}{7}+99 t^4\\
   &-1584
   t^3+9396 t^2
   -24480 t+\frac{829632}{35}\,,\\
       \tilde \cP^4_4=&-198 s^3 t-108 s^3-198 s^2 t^2+2772 s^2 t+536
   s^2+1188 s t^2-11312 s t-\frac{7968
   s}{7}\\
   &-1808 t^2+14464 t
   +\frac{56576}{35}\,,
    }
   where the other $\tilde\cP^4_i$ are given by crossing \eqref{crossP1}, while for $\tilde \cR_i^4$ we find  
   \es{quartic3}{
   \tilde \cR_1^4=&\frac{9 s^3 t}{4}-\frac{19 s^3}{4}+\frac{45 s^2
   t^2}{4}-41 s^2 t+\frac{1041 s^2}{28}+18 s
   t^3-\frac{271 s t^2}{2}+\frac{2215 s
   t}{7}\\
   &-\frac{16673 s}{70}+9 t^4-108
   t^3
   +\frac{3328 t^2}{7}-\frac{6360
   t}{7}+\frac{22569}{35}\,,\\
      \tilde \cR_2^4=&144 s^3 t-304 s^3+144 s^2 t^2-1600 s^2 t+2736
   s^2-736 s t^2+5312 s t-\frac{277792
   s}{35}\\
   &+896 t^2-5376 t
   +\frac{257664}{35}\,,\\
         \tilde \cR_3^4=&-36 s^3 t+76 s^3-108 s^2 t^2+528 s^2
   t-\frac{4476 s^2}{7}-72 s t^3+480 s
   t^2-\frac{6936 s t}{7}\\
   &+\frac{21136
   s}{35}-112 t^3
   +1008 t^2-\frac{105328
   t}{35}+\frac{104304}{35}\,.
   }
   The expressions for $\tilde \cP_i^p$ and $\tilde \cR_i^p$ for $p=1,6$ are more complicated and are relegated to Appendix \ref{R4D4R4}. 
   The large $s,t$ limit of the $p=1$ supergravity term is
      \es{largeSTPR1}{
 M_{\cP,\text{asymp}}^{1}\equiv  M^1_\cP(s,t)\underset{s,t\to\infty}\longrightarrow& 15\frac{1}{stu}\left(tuX_{12}X_{34}+stX_{13}X_{24}+suX_{14}X_{23}\right)^2\,,\\
M_{\cR,\text{asymp}}^{1}\equiv      M^1_\cR(s,t)\underset{s,t\to\infty}\longrightarrow& 12 (t-u) (Y_1 \circ Y_2) \cdot (X_3 \circ X_4) Y_{12} X_{34} \\
&      -\frac{3}{4}\frac{(t-u)^2}{s} Y_{12}^2 X_{34}^2 - 48s \left[(Y_1 \circ Y_2) \cdot (X_3 \circ X_4) \right]^2
      \,. 
   }
   For $p=4,6$, the large $s,t$ Mellin amplitudes can then be written in terms of the asymptotic $p=1$ terms as
   \es{largeSTPR}{
   M^4_\cP(s,t)\underset{s,t\to\infty}\longrightarrow& \frac{99}{15}stuM_{\cP,\text{asymp}}^{1} \,,\qquad  M^6_\cP(s,t)\underset{s,t\to\infty}\longrightarrow \frac{195}{15} stu(s^2+t^2+u^2)M_{\cP,\text{asymp}}^{1}\,,\\
      M^4_\cR(s,t)\underset{s,t\to\infty}\longrightarrow& 3stu M_{\cR,\text{asymp}}^{1}\,,\qquad
      M^6_\cR(s,t)\underset{s,t\to\infty}\longrightarrow \frac{13}{3} stu(s^2+t^2+u^2)M_{\cR,\text{asymp}}^{1}\,. 
   }
Just as we saw with the asymptotic $M_\cS^p$ in \eqref{M46}, the asymptotic $M_\cS^p$ and $M_\cR^p$ take the form of a universal polarization factor, which can be read off from the supergravity term, multiplied by the unique crossing symmetric polynomial in $s,t$ of the required degree. This is consistent with the flat space interpretation as an M-theory S-matrix with a universal polarization term multiplied by polynomials in $s,t$ for each order, as shown in \eqref{A11D} and \eqref{fR4}. Note that both $M_\cP^p$ and $M_\cR^p$ can be use in the flat space formula \eqref{FlatFromCFT}, as was originally discussed for $M_\cS^p$ in \cite{Chester:2018aca}. The numerical factors in \eqref{largeSTPR} relative to \eqref{M46} are compensated by the $\alpha$ integral in \eqref{FlatFromCFT} that depends on the dimensions of the external operators.

\subsection{From Mellin to position space}
\label{MtoP}

 To compute the integrated correlators, we need the position space expressions $\cS_i^p(U,V)$, $\cP_i^p(U,V)$, and $\cR_I^p(U,V)$ of $\tilde \cS_i^p(s,t)$, $\tilde\cP_i^p(s,t)$, and $\tilde\cR_I^p(s,t)$ for $p=4,6$. Since the latter expressions are polynomials in $s,t$, we can rewrite them as finite sums of the function $ \bar D_{r_1,r_2,r_3,r_4}(U,V)$ described in Appendix \ref{dbarApp} using its Mellin transform $ \bar D^{SSSS}_{r_1,r_2,r_3,r_4}(s,t)$ \eqref{dbar}, which is a polynomial in $s,t$ after being multiplied by an appropriate power of $U $.  For instance, for $p=4$ we find
  \es{quartic1Pos}{
 \cS^4_1=&U\left[-\frac{1536}{35} 
   \bar D_{2,1,1,2}+\frac{1728}{35} 
   \bar D_{2,1,2,3}-\frac{64}{7} 
   \bar D_{2,1,3,4}+\frac{7488}{35} 
   \bar D_{3,1,1,3}\right.\\
   &\left.-\frac{960}{7} 
   \bar D_{3,1,2,4}
   +16 
   \bar D_{3,1,3,5}
   -128 
   \bar D_{4,1,1,4}+32 
   \bar D_{4,1,2,5}+16  \bar D_{5,1,1,5}\right]\,,\\
    \cS^4_4=&U\left[\frac{1536}{35} 
   \bar D_{1,1,2,2}-\frac{6912}{35} 
   \bar D_{1,1,3,3}+\frac{512}{7} 
   \bar D_{1,1,4,4}-\frac{384}{7} 
   \bar D_{2,1,2,3}+\frac{1024}{7} 
   \bar D_{2,1,3,4}\right.\\
   &\left.-32 
   \bar D_{2,1,4,5}
   +\frac{128}{7} 
   \bar D_{3,1,2,4}-32 U \bar D_{3,1,3,5}\right]\,,
 }
and
 \es{quartic2Pos}{
 \cP^4_1=&U^2\left[\frac{41472}{35} 
   \bar D_{2,2,2,2}-\frac{13248}{7}
   \bar D_{2,2,3,3}+448
   \bar D_{2,2,4,4}-22464
   \bar D_{3,2,2,3}\right.\\
   &+20160 
   \bar D_{3,2,3,4}
   -3168 
   \bar D_{3,2,4,5}+39168 
   \bar D_{4,2,2,4}-19008 
   \bar D_{4,2,3,5}\\
   &\left.+1584 
   \bar D_{4,2,4,6}-15840 
   \bar D_{5,2,2,5}
   +3168 
   \bar D_{5,2,3,6}+1584 
   \bar D_{6,2,2,6}\right]\,,\\
    \cP^4_4=&U^2\left[-\frac{44544}{35} 
   \bar D_{2,2,2,2}+\frac{157056}{7} 
   \bar D_{2,2,3,3}-30976 
   \bar D_{2,2,4,4}+7200 
   \bar D_{2,2,5,5}\right.\\
   &+2688 
   \bar D_{3,2,2,3}
   -26240 
   \bar D_{3,2,3,4}+22176 
   \bar D_{3,2,4,5}-3168 
   \bar D_{3,2,5,6}\\
   &\left.-896 
   \bar D_{4,2,2,4}+6336 
   \bar D_{4,2,3,5}
   -3168 
   \bar D_{4,2,4,6}\right]\,,
 }
and
 \es{quartic3Pos}{
 \cR^4_1=&U^2\left[\frac{1536}{35} 
   \bar D_{2,2,1,1}-\frac{8832}{35} 
   \bar D_{2,2,2,2}+\frac{1104}{7} 
   \bar D_{2,2,3,3}-16 
   \bar D_{2,2,4,4}-\frac{6912}{7} 
   \bar D_{3,2,1,2}\right.\\
   &+\frac{13200}{7} 
   \bar D_{3,2,2,3}-608 
   \bar D_{3,2,3,4}+36 
   \bar D_{3,2,4,5}+\frac{15552}{7} 
   \bar D_{4,2,1,3}-1616 
   \bar D_{4,2,2,4}\\
&  \left. +180 U^2
   \bar D_{4,2,3,5}-1152 
   \bar D_{5,2,1,4}+288 
   \bar D_{5,2,2,5}+144 
   \bar D_{6,2,1,5}\right]\,,\\
    \cR^4_2=&U^2\left[-\frac{24576}{35}  \bar D_{2,2,2,2}+3072
   U^2 \bar D_{2,2,3,3}-1024 
   \bar D_{2,2,4,4}+3072 
   \bar D_{3,2,2,3}\right.\\
   &\left.-10240 
   \bar D_{3,2,3,4}
   +2304 U^2
   \bar D_{3,2,4,5}-1024 
   \bar D_{4,2,2,4}+2304 
   \bar D_{4,2,3,5}\right]\,,\\
    \cR^4_3=&U^2\left[-\frac{12288}{35} 
   \bar D_{2,2,1,1}+\frac{73728}{35} 
   \bar D_{2,2,2,2}-\frac{11520}{7} 
   \bar D_{2,2,3,3}+256 
   \bar D_{2,2,4,4}\right.\\
   &+\frac{239616}{35} 
   \bar D_{3,2,1,2}
   -\frac{103680}{7} 
   \bar D_{3,2,2,3}+6144 
   \bar D_{3,2,3,4}-576 
   \bar D_{3,2,4,5}\\
   &\left.-9216 
   \bar D_{4,2,1,3}+9984 
   \bar D_{4,2,2,4}
   -1728 
   \bar D_{4,2,3,5}+2048 
   \bar D_{5,2,1,4}-1152 
   \bar D_{5,2,2,5}\right]\,,
     }
     where the other $\cS_i^4$ and $\cP_i^4$ are related by crossing \eqref{Crossing} using \eqref{crossDbar}. 
The analogous expressions for $p=6$ are given in Appendix \ref{R4D4R4}\@. Note that the $p=1$ Mellin amplitude is an infinite sum of poles in $s,t$ and so would correspond to an infinite sum of $ \bar D$'s, which is why we do not consider this term in the rest of the paper.

\subsection{Integrals of $\langle SSSS\rangle $, $\langle PPPP\rangle $, and $\langle SSPP\rangle $}
\label{ints}

We now compute the $S^3$ integrals of the $c_T^{-\frac53}$ and $c_T^{-\frac{19}{9}}$ terms in $\langle SSSS\rangle $, $\langle PPPP\rangle $, and $\langle SSPP\rangle $ in position space, using the expressions derived in the previous section. Recall that in Section \ref{S3} we reduced the 12-dimensional integral over each 3-component $\vec x_i$ to a 2-dimensional integral over $r$ and $\theta$, where $U=1+r^2-2r\cos\theta$ and $V=r^2$, and the measure was proportional to $U^{-1}\bar D_{1,1,1,1}(U,V)$, $U^{-2}\bar D_{2,2,2,2}(U,V)$, and $U^{-2}\bar D_{1,1,2,2}(U,V)$ for $\langle SSSS\rangle $, $\langle PPPP\rangle $, and $\langle SSPP\rangle $. The powers of $U$ in the former expressions exactly cancel those in the latter, as given in Section \ref{MtoP}, so that our integral is over a sum of pairs of $\bar D$ functions. We can then plug in the explicit expressions for the $\bar D$ functions using the algorithm in Appendix \ref{dbarApp}, perform the integral numerically to high precision for each component $\cS_i^p$, $\cP_i^p$, and $\cR_I^p$, and assemble these ingredients into the formula \eqref{d4Fdm4} for derivatives of the free energy. We find that our numerical results are consistent with the analytic expressions
\es{intResults}{
-\frac{1}{c_T^2}\frac{\partial^4F}{\partial m_a^4}&=\frac{1}{c_T^{\frac53}}B_4^4\frac{3\pi^4}{560}+\frac{1}{c_T^{\frac{19}{9}}}\left[B_6^4\frac{3\pi^4}{560}-B_6^6\frac{27\pi^4}{616}\right]\,,\\
-\frac{1}{c_T^2}\frac{\partial^4F}{\partial m_1^2\partial m_2^2}&=\frac{1}{c_T^{\frac53}}B_4^4\frac{\pi^4}{1120}+\frac{1}{c_T^{\frac{19}{9}}}\left[B_6^4\frac{\pi^4}{1120}-B_6^6\frac{13\pi^4}{1848}\right]\,.\\
}


\subsection{Localization for $F(m_1,m_2)$ in ABJM}
\label{localization}

So far the discussion has applied to any $\mathcal{N}=8$ SCFT with a large $c_T$ expansion. We will now specify to ABJM$_{N,k}$ for $k=1,2$ using the all orders in $1/c_T$ results for $F(m_1,m_2)$ for this theory. 

For ABJM$_{N,k}$, the mass deformed partition function has been computed from localization \cite{Kapustin:2009kz}, and takes the form
 \es{ZABJM}{
 & Z_{\text{ABJM}_{N,k}} = \frac{1}{N!^2} \int d^N \lambda\, d^N \mu\, e^{i \pi k \left[ \sum_i \lambda_i^2 - \sum_j \mu_j^2 \right]} \\
  &\times \frac{\prod_{i<j} \left( 4 \sinh^2 \left[ \pi (\lambda_i - \lambda_j) \right] \right) 
    \prod_{i<j} \left( 4 \sinh^2 \left[ \pi (\mu_i - \mu_j) \right] \right) }{\prod_{i, j} \left(
     4 \cosh \left[\pi (\lambda_i - \mu_j + m_1/2 + m_2/2 ) \right] 
     \cosh \left[\pi (\mu_i - \lambda_j + m_1/2 - m_2/2 ) \right]  \right) } \,.
 }
Using the Fermi gas technique \cite{Marino:2011eh,Nosaka:2015iiw}, this quantity was computed to all orders in $1/N$:
 \es{GotZABJM}{
 & Z_\text{ABJM$_{N,k}$} \approx e^A C^{-\frac 13} \text{Ai}\left[C^{-\frac 13} (N-B) \right] \,,\\
  C &= \frac{2}{\pi^2 k (1 + m_+^2) (1 + m_-^2)} \,, \qquad
   B = \frac{\pi^2 C}{3} - \frac{1}{6k} \left[ \frac{1}{1 + m_+^2} + \frac{1}{1 + m_-^2} \right] + \frac{k}{24} \,, \\
  A&= \frac{{\cal A}[k(1 + i m_+)] + {\cal A}[k(1 - i m_+)] +  {\cal A}[k(1 + i m_-)] + {\cal A}[k(1 - i m_-)] }{4}  \,,
 } 
 where $m_\pm \equiv m_1 \pm m_2$, and the function ${\cal A}$ is given by
\es{constantMap}{
{\cal A}(k)=\frac{2\zeta(3)}{\pi^2k}\left(1-\frac{k^3}{16}\right)+\frac{k^2}{\pi^2}\int_0^\infty dx\frac{x}{e^{kx}-1}\log\left(1-e^{-2x}\right)\,,
}
with derivatives 
\es{Aprime}{
{\cal A}''(1)&=\frac16+\frac{\pi^2}{32}\,,\qquad\qquad \,\, \,\,{\cal A}''(2)=\frac{1}{24}\,,\\
{\cal A}''''(1)&=1+\frac{4\pi^2}{5}-\frac{\pi^4}{32}\,,\qquad {\cal A}''''(2)=\frac{1}{16}+\frac{\pi^2}{80}\,.
} 
Using \eqref{ZABJM}, \eqref{cTfromF}, and \eqref{d4Fdm4} we can expand both $c_T$ and $\frac{\partial^4F(m)}{\partial m_a^2\partial m_b^2}$ to all orders in $1/N$ in ABJM$_{N,k}$ for $k=1,2$, and then write $\frac{\partial^4F(m)}{\partial m_a^2\partial m_b^2}$ for $a,b=1,2$ as an all orders expansion in $1/c_T$. For $k=2$ we find
\es{mass4ABJMcT}{
-\frac{1}{c_T^2}\frac{\partial^4 F^\text{ABJM$_{N,2}$}}{\partial m_a^4}&=\left({12\pi^4}\right)^{\frac13}\frac{1}{c_T^{\frac53}}+\left(\frac{\pi^2}{5}-5\right)\frac{1}{c_T^{2}}-28\left(\frac{\pi^2}{12}\right)^{\frac13}\frac{1}{c_T^{\frac73}}+O\left(c_T^{-\frac83}\right)\,,\\
-\frac{1}{c_T^2}\frac{\partial^4 F^\text{ABJM$_{N,2}$}}{\partial m_1^2\partial m_2^2}&=\frac{\pi^2}{8}\frac{1}{c_T}+\left(\frac{\pi^4}{18}\right)^{\frac13}\frac{1}{c_T^{\frac53}}+\left(\frac{\pi^2}{5}-\frac53\right)\frac{1}{c_T^{2}}-\frac{20}{3}\left(\frac{\pi^2}{12}\right)^{\frac13}\frac{1}{c_T^{\frac73}}+O\left({c_T^{-\frac{8}{3}}}\right)\,.
} 
For $k=1$, recall that that ABJM$_{N,1}$ is a product between the free theory ABJM$_{1,1}$ and an interacting theory ABJM$_{N,1}^\text{int}$. Recall from Section \ref{MSQUARED} that $c_T^\text{free}=16$ and $-\frac{\partial^4F^{\text{free}}}{\partial m_a^2\partial m_b^2}=\frac{\pi^4}{4}$. For ABJM$_{N,1}^\text{int}$, we should then expand derivatives of $F^{\text{ABJM$_{N,1}^\text{int}$}}\equiv F^{\text{ABJM$_{N,1}$}}-F^{\text{free}}$ in terms of $c_T^{\text{ABJM$_{N,1}^\text{int}$}}\equiv c_T^{\text{ABJM$_{N,1}$}}-c_T^{\text{free}}$, for which we find
\es{mass4ABJMcTint}{
-\frac{1}{c_T^2}\frac{\partial^4 F^{\text{ABJM$_{N,1}^\text{int}$}}}{\partial m_a^4}&=\left({48\pi^4}\right)^{\frac13}\frac{1}{c_T^{\frac53}}+\left(-\frac{\pi^4}{32}+\frac{4\pi^2}{5}-5\right)\frac{1}{c_T^{2}}-28\left(\frac{4\pi^2}{3}\right)^{\frac13}\frac{1}{c_T^{\frac73}}+O\left(c_T^{-\frac83}\right)\,,\\
-\frac{1}{c_T^2}\frac{\partial^4 F^{\text{ABJM$_{N,1}^\text{int}$}}}{\partial m_1^2\partial m_2^2}&=\frac{\pi^2}{8}\frac{1}{c_T}+\left(\frac{2\pi^4}{9}\right)^{\frac13}\frac{1}{c_T^{\frac53}}+\left(-\frac{\pi^4}{32}+\frac{107\pi^2}{40}-\frac53\right)\frac{1}{c_T^{2}}-\frac{20}{3}\left(\frac{4\pi^2}{3}\right)^{\frac13}\frac{1}{c_T^{\frac73}}+O\left({c_T^{-\frac{8}{3}}}\right)\,.
} 
Comparing \eqref{mass4ABJMcT} to \eqref{mass4ABJMcTint}, we find that the tree level terms of order $c_T^{-n}$ (namely $c_T^{-1}$, $c_T^{-\frac 53}$, and $c_T^{-\frac 73}$ in these expressions) come with $k$-dependence of the form $k^{1-n} c_T^{-n}$. This can be argued as follows.  When we compactify 11d M-theory on $AdS_4\times S^7/\mathbb{Z}_k$, each term in the AdS$_4$ effective action is proportional to $\Vol(S^7 / \Z_k) \propto 1/k$ times an appropriate power of $ \ell_p/L$ determined by dimensional analysis.  Since $ \ell_p/L \propto (k c_T)^{-1/9}$ (see \eqref{cTAdS}), it follows that the terms in the free energy indeed scale as $k^{-1} (k c_T)^{-n}$.  If we multiply the free energy by $1/c_T^2 = k^2 / (k c_T)^2$, as in \eqref{mass4ABJMcT} and \eqref{mass4ABJMcTint}, then each tree level term will be proportional to $k$ times a power of $k c_T$, as observed from \eqref{mass4ABJMcT} and \eqref{mass4ABJMcTint}.  No similar pattern exists for the 1-loop term $c_T^{-2}$, which comes from 1-loop in 11d that has a complicated dependence on $k$.

\subsection{Deriving $B_4^4,B_6^6,B_4^6$}
\label{final}

We can now plug these localization values into the two constraints \eqref{d4Fdm4} and compare to \eqref{intResults} to extract the coefficient $B_4^4$ and $B_6^6,B_4^6$, for the order $c_T^{-\frac53}$ and $c_T^{-\frac{19}{9}}$ Mellin amplitudes, respectively. For $c_T^{-\frac53}$, we find that both constraints are satisfied by the same value
\es{R4final}{
B_4^4=(6\pi)^{\frac13}\frac{1120}{3\pi^3k^{\frac23}}\,.
}
This matches the value previously derived in \cite{Chester:2018aca} using the 1d sector of ABJM$_{N,k}$, which was shown to recover the known coefficient of the $R^4$ term in 11d M-theory. This precise match is a very nontrivial check on our formalism. 

For $c_T^{-\frac{19}{9}}$, we note that such a coefficient does not appear in the localization results \eqref{mass4ABJMcT}, so that from \eqref{intResults} we find
\es{{det}}{
0=B_6^4\frac{3\pi^4}{560}-B_6^6\frac{27\pi^4}{616}\,,\qquad0=B_6^4\frac{\pi^4}{1120}-B_6^6\frac{13\pi^4}{1848}\,.\\
}
The matrix formed by these constraints has determinant $\pi^8/689920\neq0$, so these constraints are linearly independent, which implies that
\es{D4R4final}{
B_6^6=B_4^6=0\,.
}
In the flat space limit, this implies that the $D^4R^4$ interaction in 11d M-theory is absent, as was previously argued using purely string theory reasoning.

\section{Discussion}
\label{DISCUSSION}

The main result of this work is the derivation of the absence of the protected $D^4R^4$ term in the 11d M-theory S-matrix, which had previously been derived using duality arguments between M-theory and string theory, by showing that the corresponding $c_T^{-\frac{19}{9}}$ term is absent in the large $c_T$ expansion of the dual 4-point function in the 3d CFT called ABJM. This is the first check of AdS/CFT to this order. The dual 4-point function was fixed to this order by supersymmetry as well as two constraints coming from the two linearly independent quartic mass terms in the mass deformed $S^3$ free energy $F(m_a)$, which can be computed to all orders in $1/c_T$ using localization \cite{Kapustin:2009kz} and the Fermi gas formalism \cite{Marino:2011eh,Nosaka:2015iiw}. A nontrivial check on these constraints was the recovery of the $R^4$ term, which had previously been computed in \cite{Chester:2018aca} and matched to M-theory using $F(m_a)$ and the 4-point function in the 1d topological sector. As part of the derivation, our main technical innovations are explicit relations between 4-point functions of many of the operators in the stress tensor multiplet, as well as a compact formula for the integral of any CFT 4-point function over $S^d$ in general dimension $d$.

There is one more protected term in the M-theory S-matrix: $D^6R^4$. This term was also computed using duality arguments and string theory, and corresponds to the $c_T^{-\frac73}$ coefficient in the dual CFT 4-point function. To fix this term on the CFT side we need a new constraint on the 4-point function in addition to the two from $F(m_a)$. This constraint could come from considering the free energy on the squashed sphere $F(b)$, where $n$ derivatives of the squashing parameter $b$ give integrated $n$-point functions of operators in the stress tensor multiplet. For ABJM, $F(b)$ has been computed as an $N$-dimensional integral using localization \cite{Hama:2011ea,Imamura:2011wg }. Unlike $F(m_a)$, however, no all orders in $1/N$ (and thus $1/c_T$) result is known yet for $F(b)$.\footnote{Except for specific values of $b$ \cite{Hatsuda:2016uqa}.} It would be interesting to derive such a formula using the Fermi gas formalism, which would then allow us to derive $D^6R^4$ from ABJM using the methods in this paper.

The idea in this paper of deriving constraints on 4-point functions from the mass deformed sphere free energy has applications to theories in other dimensions and/or with less supersymmetry.   In $d=3$, the all orders in $1/c_T$ formula for the ABJM $S^3$ free energy deformed by two masses can also be applied to $\mathcal{N}=6$ ABJM with gauge group $U(N)_k\times U(N)_{-k}$ for $k>2$, which in the large $N,k$ limit is dual to a Type IIA string theory background. In $d=4$, the $S^4$ free energy deformed by one mass was computed by localization \cite{Pestun:2007rz} for $\mathcal{N}=4$ super-Yang-Mills (SYM), whose large $N$ and large 't Hooft coupling limit has a dual in Type IIB string theory. In $d=5$, the $S^5$ free energy deformed by one mass has also been computed \cite{Imamura:2012xg,Imamura:2012bm,Chang:2017cdx} for various 5d SCFTs with holographic duals. In all these cases, we can use the mass deformed free energy to fix a CFT 4-point function to some order in the large $N$ expansion, which could be used to derive the dual quantum gravity S-matrix to the same order.

Ultimately, since the mass deformed free energy is a protected quantity, we can expect that it can be used to derive only the protected terms in the corresponding S-matrix. To explore the unprotected terms, we will need to derive unprotected CFT data in these holographic theories. The only known method of this sort is the numerical conformal bootstrap, which has been applied to the $d=3,4,5$ holographic theories mentioned above in \cite{Chester:2014fya,Agmon:2017xes,Beem:2013qxa,Beem:2016wfs,Chang:2017cdx}. We hope that as the precision of the numerical bootstrap studies increases, it will eventually become feasible to derive the full quantum gravity S-matrix from CFT\@.

\section*{Acknowledgments} 

We thank Nathan Agmon, Ofer Aharony, Igor Klebanov, Yifan Wang, and Xi Yin for useful discussions.  DJB, SMC, and SSP, are supported in part by the Simons Foundation Grant No~488651. DJB is also supported in part by the General Sir John Monash Foundation. SMC is also supported in part by the Bershadsky Family Scholarship in Science or Engineering.  

\appendix

\section{The $E$ invariants}
\label{EMatrices}

We use the following $E_{aIA}$ symbols, written as $8 \times 8$ matrices for given $a$:
 \es{E1}{
  E_1 &= \begin{pmatrix}
   I & 0 & 0 & 0 \\
   0 & I & 0 & 0 \\
   0 & 0 & I & 0 \\
   0 & 0 & 0 & I
  \end{pmatrix} \,, 
 \quad
  E_2 = \begin{pmatrix}
   - i \sigma_2 & 0 & 0 & 0 \\
   0 & -i \sigma_2 & 0 & 0 \\
   0 & 0 & -i \sigma_2 & 0 \\
   0 & 0 & 0 & i \sigma_2 
  \end{pmatrix} \,, \quad
   E_3 = \begin{pmatrix}
    0 & -\sigma_3 & 0 & 0 \\
    \sigma_3 & 0 & 0 & 0 \\
    0 & 0 & 0 & -I \\
    0 & 0 & I & 0 
   \end{pmatrix} \,,\\
   E_4&= \begin{pmatrix}
   0 & -\sigma_1 & 0 & 0 \\
   \sigma_1 & 0 & 0 & 0 \\
   0 & 0 & 0 & -i \sigma_2 \\
   0 & 0 & -i\sigma_2 & 0
  \end{pmatrix} \,, 
 \quad
  E_5 = \begin{pmatrix}
   0 & 0 & -\sigma_3 & 0 \\
   0 & 0 & 0 & I \\
   \sigma_3 & 0 & 0 & 0 \\
   0 & -I & 0 & 0
  \end{pmatrix} \,, \quad
   E_6 = \begin{pmatrix}
    0 & 0 & -\sigma_1 & 0 \\
    0 & 0 & 0 & i \sigma_2 \\
    \sigma_1 & 0 & 0 & 0 \\
    0 & i \sigma_2 & 0 & 0 
   \end{pmatrix} \,, \\
   E_7&= \begin{pmatrix}
   0 & 0 & 0 & -I \\
   0 & 0 & -\sigma_3 & 0 \\
   0 & \sigma_3 & 0 & 0 \\
   I & 0 & 0 & 0
  \end{pmatrix} \,, 
 \quad
  E_8 = \begin{pmatrix}
   0 & 0 & 0 & -i \sigma_2 \\
   0 & 0 & -\sigma_1 & 0 \\
   0 & \sigma_1 & 0 & 0 \\
   -i \sigma_2 & 0 & 0 & 0 
  \end{pmatrix} \,, 
 }
where $I$ is the $2 \times 2$ identity matrix and $0$ is interpreted as a $2\times 2$ matrix with vanishing entries.  From these matrices, one can construct the $\mf{so}(8)$ gamma matrices as
 \es{GammaMatrices}{
  \Gamma_a = \begin{pmatrix} 0 & E_a \\
  E_a^T & 0 
  \end{pmatrix} \,.
 }
The Clifford algebra $\{\Gamma_a, \Gamma_b\} =2 \delta_{ab}$ is equivalent to $E_a E_b^T + E_b E_a^T = 2 \delta_{ab}$.

\section{Ward Identities}
\label{WardAppendix}

\subsection{Structures for $\chi$ and $j^\mu$}
We will first expand correlators containing $\chi$ in the allowed $SO(8)_R$ and conformally invariant structures, which are derived in \cite{Iliesiu:2015akf}. Using the notation $\vec x_{ij} = \vec x_i - \vec x_j$, $x_{ij} = |\vec x_{ij}|$ and $\slashed x_{ij} = (x_i^\m-x_j^\m)\sigma_\m^{\alpha\beta}$, we normalize $\chi$ so that the two point function is given by:
\begin{equation}\langle \chi^\alpha(\vec x_1,X_1,Y_1)\chi^\beta(\vec x_2,X_2,Y_2)\rangle =\frac{i\slashed x_{12}}{x_{12}^4}.\end{equation}
We expand the four point function as:
\begin{equation}\begin{split}
\langle S(\vec x_1,Y_1)&S(\vec x_2,Y_2)\chi^\alpha(\vec x_3,X_3,Y_3)\chi^\beta(\vec x_4,X_4,Y_4)\rangle \\
= &\frac{i\slashed x_{34}^{\alpha\beta}}{x_{12}^2x_{34}^4}\biggr[Y_{12}X_{34}(\cA_{11}Y_{12}Y_{34} + \cA_{12}Y_{13}Y_{24}+\cA_{13} Y_{14}Y_{23}) \\
& + (Y_1\circ Y_2)\cdot(X_3\circ X_4)(\cA_{14}Y_{12}Y_{34} + \cA_{15}Y_{13}Y_{24}+\cA_{16} Y_{14}Y_{23})\biggr] \\
&+\frac{i(\slashed x_{13}\slashed x_{24}\slashed x_{12})^{\alpha\beta}}{2x_{12}^4x_{34}^4}\biggr[Y_{12}X_{34}(\cA_{21}Y_{12}Y_{34} + \cA_{22}Y_{13}Y_{24}+\cA_{23} Y_{14}Y_{23}) \\
& + (Y_1\circ Y_2)\cdot(X_3\circ X_4)(\cA_{24}Y_{12}Y_{34} + \cA_{25}Y_{13}Y_{24}+\cA_{26} Y_{14}Y_{23})\biggr],
\end{split}\end{equation}
\begin{equation}\begin{split}
\langle P(\vec x_1,X_1)&P(\vec x_2,X_2)\chi^\alpha(\vec x_3,X_3,Y_3)\chi^\beta(\vec x_4,X_4,Y_4)\rangle \\
= &\frac{i\slashed x_{34}^{\alpha\beta}}{x_{12}^4x_{34}^4}\biggr[X_{12}Y_{34}(\cC_{11}X_{12}X_{34} + \cC_{12}X_{13}X_{24}+\cC_{13} X_{14}X_{24}) \\
& + (X_1\circ X_2)\cdot(Y_3\circ Y_4)(\cC_{14}X_{12}X_{34} + \cC_{15}X_{13}X_{24}+\cC_{16} X_{14}X_{24})\biggr] \\
&+\frac{i(\slashed x_{13}\slashed x_{24}\slashed x_{12})^{\alpha\beta}}{2x_{12}^6x_{34}^4}\biggr[X_{12}Y_{34}(\cC_{21}X_{12}X_{34} + \cC_{22}X_{13}X_{24}+\cC_{23} X_{14}X_{24}) \\
& + (X_1\circ X_2)\cdot(Y_3\circ Y_4)(\cC_{24}X_{12}X_{34} + \cC_{25}X_{13}X_{24}+\cC_{26} X_{14}X_{24})\biggr] \,, 
\end{split}\end{equation}
along with:
\begin{equation}\begin{split}
\langle S(\vec x_1,Y_1)&P(\vec x_2,Y_2)\chi^\alpha(\vec x_3,X_3,Y_3)\chi^\beta(\vec x_4,X_4,Y_4)\rangle \\
&= -\frac{\slashed x_{13}\slashed x_{14}}{2x_{12}^4x_{34}^4}\biggr[Y_{13}X_{24}(Y_{14}X_{23}\cB_{11} + (Y_1\circ Y_4)\cdot(X_2\circ X_3)\cB_{12}) \\
&+ (Y_1\circ Y_3)\cdot(X_2\circ X_4)(Y_{14}X_{23}\cB_{13} + (Y_1\circ Y_4)\cdot(X_2\circ X_3)\cB_{14}) \biggr] \\
&- \frac{(\slashed x_{23}\slashed x_{24})x_{14}^2}{2x_{12}^4x_{34}^4x_{24}^2}\biggr[Y_{13}X_{24}(Y_{14}X_{23}\cB_{21} + (Y_1\circ Y_4)\cdot(X_2\circ X_3)\cB_{22}) \\
&+ (Y_1\circ Y_3)\cdot(X_2\circ X_4)(Y_{14}X_{23}\cB_{23} + (Y_1\circ Y_4)\cdot(X_2\circ X_3)\cB_{24}) \biggr] \,.
\end{split}\end{equation}

For correlators involving the $R$--symmetry current $j^\mu$, spacetime structures were computed using the embedding space formalism. Since $j^\mu$ is a conserved current there is actually only one allowed conformal structure; we however did not impose this condition. We could then test whether our final answers satisfied $\partial^\m j_\m = 0$, serving as a non-trivial check on the Ward identities.

Our normalization of $j^\mu$ is such that:
\begin{equation}\langle j^\mu(\vec x_1,Y_1,Y_1')j^\nu(\vec x_2,Y_2,Y_2')\rangle = \frac{4[(Y_1\cdot Y_2)(Y_1'\cdot Y_2')-(Y_1\cdot Y_2')(Y_1'\cdot Y_2)]}{x_{12}^4} \left(g^{\mu\nu}-2\frac{x_{12}^\mu x_{12}^\nu}{x_{12}^2}\right)\,.\end{equation}
We then expand the four point functions of scalars and currents as:
\begin{equation}\begin{split}\langle S(\vec x_1,Y_1)&S(\vec x_2,Y_2)S(\vec x_3,Y_3)j^\m(\vec x_4,Y_4,Y_5)\rangle \\
&= \frac 1 {x_{12}^2x_{34}^2}\left(\frac{x_{24}^\m}{x_{24}^2}-\frac{x_{34}^\m}{x_{34}^2}\right)\biggr[(Y_4\circ Y_5)\cdot(Y_1\circ Y_2) Y_{13}Y_{23} \cW_{11} \\
&+ (Y_4\circ Y_5)\cdot(Y_2\circ Y_3) Y_{13}Y_{12} \cW_{12}+(Y_4\circ Y_5)\cdot(Y_3\circ Y_1) Y_{12}Y_{23} \cW_{13}\biggr] \\
&+\frac 1 {x_{12}^2x_{34}^2}\left(\frac{x_{24}^\m}{x_{24}^2}-\frac{x_{14}^\m}{x_{14}^2}\right)\biggr[(Y_4\circ Y_5)\cdot(Y_1\circ Y_2) Y_{13}Y_{23} \cW_{21} \\
&+ (Y_4\circ Y_5)\cdot(Y_2\circ Y_3) Y_{13}Y_{12} \cW_{22}+(Y_4\circ Y_5)\cdot(Y_3\circ Y_1) Y_{12}Y_{23} \cW_{23}\biggr] \,, \\
\langle P(\vec x_1,X_1)&P(\vec x_2,X_2)S(\vec x_3,Y_3)j^\m(\vec x_4,Y_4,Y_5)\rangle \\
&= (X_1\circ X_2)\cdot[(Y_3\circ Y_4)Y_{35} - (Y_3\circ Y_5)Y_{34}]X_{12} \\
&\biggr(\frac 1 {x_{12}^4x_{34}^2}\left(\frac{x_{24}^\m}{x_{24}^2}-\frac{x_{34}^\m}{x_{34}^2}\right)  \mathcal Y_1 
+ \frac 1 {x_{12}^4x_{34}^2}\left(\frac{x_{24}^\m}{x_{24}^2}-\frac{x_{14}^\m}{x_{14}^2}\right) \mathcal Y_2\biggr) \,, \\
\langle S(\vec x_1,Y_1)&S(\vec x_2,Y_2)P(\vec x_3,X_3)j^\m(\vec x_4,Y_4,Y_5)\rangle \\
&= \frac{x_{14}^2 (\vec x_{24}\times \vec x_{34}) + x_{24}^2(\vec x_{34}\times\vec  x_{14}) + x_{34}^2(\vec x_{14}\times\vec  x_{24}))}{x_{12}^4x_{34}^6}\\
&\times (Y_1\circ Y_2)\cdot[(X_3\circ X_4)X_{35} - (X_3\circ X_5)X_{34}]Y_{12} \mathcal X_1 \,.
\end{split}\end{equation}

\subsection{Ward Identities for Scalar Correlators}
We will now give explicit expressions for the differential operators in \eqref{wards}. For the correlator $\langle SSPP\rangle$, the three equations are
\es{SSPP1}{
\cR_1(U,V) &= \frac 1 4\biggr[4+\left(U^2 - 4U\right)\partial_U + \left(4 + U - 2U^2 + 7 U V - 4V^2\right)\partial_V  \\
& \ \ \ +2U(2V-U+2)(U\partial_U^2+(U+V-1)\partial_U\partial_V+V\partial_V^2) \biggr]\cS_1(U,V) \,, \\
\cR_2(U,V) &= 16U^2\left[\partial_U+\frac{1+2U-V} U\partial_V+2U\partial_U^2+(2U+2V-2)\partial_U\partial_V+2U\partial_V^2\right]\cS_1(U,V) \,, \\
\cR_3(U,V) &= 4\bigg[U^2\partial_U-(2 + U - 4 V - 3 U V + 2 V^2)\partial_V\\
& \ \ \ + 2U (V-1)(U\partial_U^2+(U+V-1)\partial_U\partial_V+U\partial_V^2)\bigg]\cS_1(U,V) \,.
}
For $\langle PPPP\rangle$ we will give expressions for $\cP_1$ and $\cP_4$. The other expressions can be derived from this by applying the crossing relations \eqref{Crossing}.
 \es{P1}{
\cP_1&(U,V) = \bigg[1 + (U + 4 U^2 + 4 (V-1)^2 - 7 U V) \partial_V + (U-1) U \partial_U +  U^2 (9U + 8 V-3) \partial_U^2\\
&+ (28 U^2 + 9U - 15UV + 4 (V-1)^2) V \partial_V^2 + 2 U (6 U^2 -4U - 2 (V-1)^2 + 13UV) \partial_U\partial_V\\
& + 
 4 U ( 6 U - V + 1) V^2 \partial_V^3 + 
 4 U^3 (U + 4 V-1) \partial_U^3 +
 4 U (7 U^2 - 6U + 9UV + (V-1)^2) V \partial_U\partial_V^2\\
&+ 
 4 U^2 (1 + U^2 - 2U - 4 V + 3 V^2 + 11 UV)) \partial_U^2\partial_V + 4 U^2 V^3 \partial_V^4 + 
 8 U^2 V^2 (U + V-1) \partial_U\partial_V^3 \\
& + 
 4 U^2 V (U^2 -2U+4UV+ (V-1)^2 ) \partial_U^2\partial_V^2  + 
 8 U^3 V ( U + V - 1) \partial_U^3\partial_V + 4 U^4 V \partial_U^4\bigg]\cS_1(U,V) .
 }
 \es{P4}{
\cP_4&(U,V) = \frac U {4V}(U - V - 1)\cS_4(U,V) \\
&+\frac U{2V^2} \biggr[
(-1 + U + 3 U^2 - 5 U^3 + 2 U^4 + 2 V - U V + 14 U^2 V - 15 U^3 V + 
 11 U V^2 + 31 U^2 V^2 \\
 &- 2 V^3 - 19 U V^3 + V^4) \partial_V + U (-1 + 3 U - 3 U^2 + U^3 - V + 8 U V - 7 U^2 V + V^2 + 17 U V^2\\
 & - 11 V^3) \partial_U +  2 U V (-1 + 3 U - 3 U^2 + U^3 + 3 V - 20 U V + 17 U^2 V + 5 V^2 - 7 U V^2 - 
 11 V^3) \partial_V^2 \\
 &+ 
 2 U (1 - 4 U + 6 U^2 - 4 U^3 + U^4 - V - 8 U V + 19 U^2 V - 10 U^3 V + 
 3 V^2 - 40 U V^2 + 67 U^2 V^2\\
 & + 7 V^3 - 48 U V^3 - 10 V^4) \partial_U\partial_V + 
 2 U^2 (-1 + 3 U - 3 U^2 + U^3 - 4 V + 14 U V - 10 U^2 V - 11 V^2\\
 & + 
 47 U V^2 - 38 V^3) \partial_U^2 +  4 U V^3 (1 + 12 U^2 - V^2 - 13U + 11 UV) \partial_V^3 + 
 4 U V^2 (-1 + 13 U - 23 U^2 \\
 &+ 11 U^3 + V - 11 U V + 16 U^2 V + V^2 -  26 U V^2 - V^3) \partial_U\partial_V^2 - 
 4 U^2 V (-1 + 3 U - 3 U^2 + U^3\\
 & - 9 V + 35 U V - 26 U^2 V - 5 V^2 + 10 U V^2 + 
 15 V^3) \partial_U^2\partial_V - 
 4 U^3 V (1 + U^2 + 9 V + 14 V^2 - 2U\\
 & - 15UV) \partial_U^3 +  8 U^2 (-1 + U - V) V^4 \partial_V^4 +  16 U^2 V^3 (1 - 2 U + U^2 - V^2) \partial_U\partial_V^3 + 
 8 U^2 V^2 (-1 + 3 U\\
 & - 3 U^2 + U^3 + V - 4 U V + 3 U^2 V + V^2 - 3 U V^2 - V^3) \partial_U^2\partial_V^2 + 
 16 U^3 V^2 (1 - 2 U + U^2 - V^2) \partial_U^3\partial_V \\
 &+ 8 U^4 (-1 + U - V) V^2 \partial_U^4
\biggr]\cS_1(U,V) \\
&-\frac 1 {4V}\biggr[2U - 2U^2 - 2V + 2V^2+(2 U - 4 U^2 + 2 U^3 + 4 V + 4 U V + 4 U^2 V - 8 V^2 - 10 U V^2\\
& + 4 V^3)\partial_V + (-2 U^2 + 2 U^3 + 2 U V - 2 U V^2)\partial_U+(-4 U V + 8 U^2 V - 4 U^3 V + 8 U V^2 + 8 U^2 V^2\\
& - 4 U V^3)(\partial_V^2+\partial_U\partial_V)\biggr]\cS_2(U,V) \\
&+\frac 1 {2V^2}\biggr[1 - 3 U + 3 U^2 - U^3 - 3 V + 3 U^2 V + 2 V^2 - 2 U V^2+(2 V - 3 U V - 3 U^2 V + 7 U^3 V\\
& - 3 U^4 V - 4 V^2 - U V^2 - 8 U^2 V^2 + 13 U^3 V^2 + 2 V^3 - 5 U V^3 - 17 U^2 V^3 + 7 U V^4)\partial_V+(-U + 4 U^2\\
& - 6 U^3 + 4 U^4 - U^5 + 3 U V + U^2 V - 11 U^3 V + 
7 U^4 V - 2 U V^2 + U^2 V^2 - 11 U^3 V^2 + 5 U^2 V^3)\partial_U\\
&+(2 U V^2 - 6 U^2 V^2 + 6 U^3 V^2 - 2 U^4 V^2 - 2 U V^3 - 4 U^2 V^3 + 
6 U^3 V^3 - 2 U V^4 - 6 U^2 V^4 + 2 U V^5)\partial_V^2\\
&+(-2 U V + 8 U^2 V - 12 U^3 V + 8 U^4 V - 2 U^5 V + 4 U V^2 - 
2 U^2 V^2 - 8 U^3 V^2 + 6 U^4 V^2 - 2 U V^3\\
& - 6 U^3 V^3 + 2 U^2 V^4)\partial_U\partial_V\biggr]\cS_3(U,V) \,.
}

The expressions \eqref{SSPP1}--\eqref{P4} assume that the operators $S(\vec{x}, Y)$ and $P(\vec{x}, X)$ are normalized such that their two-point functions are
\es{TwoPoint}{
  \langle S(\vec{x}_1, Y_1) S(\vec{x}_2, Y_2) \rangle 
   = \frac{Y_{12}^2}{x_{12}^2} \,, \qquad
    \langle P(\vec{x}_1, X_1) P(\vec{x}_2, X_2) \rangle 
   = \frac{X_{12}^2}{x_{12}^4} \,.
 } 
This means that in the small $U$ limit, the functions $\cS_1$, $\cP_1$, $\cR_1$ approach $1$ as $U \to 0$.

\subsection{Ward Identities from $\langle SSS\chi\rangle$}
By considering the supersymmetric variation $\delta\langle SSS\chi\rangle$, we can compute $\langle SS\chi\chi\rangle$ and $\langle SSSj\rangle$ in terms of $\langle SSSS\rangle$. These expressions are first order differential operators and are relatively simple, so we will give them explicitly here as an examples. Other correlators can be found in the attached \verb|Mathematica| file.

First we will give the $\langle SS\chi\chi\rangle$ expressions:
\begin{equation}\begin{split}
\cA_{11} &= [2-2 U \partial_U - U\partial_V] \cS_1 \,,\\
\cA_{12} &= \left[ - \frac{2 + U - 2 V} U  + (2 + U - 2 V) \partial_U + (U - 2 V)\partial_V\right]\cS_2 + \cS_6 \,,\\
\cA_{13} &= \left[\frac{(2 + U - 2 V)} U + \frac{V (-2 + U + 2 V)} U \partial_U + 2 V \partial_V\right]\cS_3+ \cS_5 \,, \\
\cA_{14} &= 8U\partial_V \cS_1 \,,\\
\cA_{15} &= 8[1- U\partial_U - U\partial_V] \cS_2 \,,\\
\cA_{16} &= 8[V\partial_U-1] \cS_3 ,\\
\cA_{21} &= [U^2\partial_U + U (1 + V)\partial_V] \cS_1 \,,\\
\cA_{22} &= \left[2-2U\partial_U-U\partial_V\right]\cS_2 \,,\\
\cA_{23} &= -\left[2(U-2)V\partial_U + UV\partial_V\right]\cS_3  \,,\\
\cA_{24} &= [8 U^2\partial_U + 8U (V-1)\partial_V]\cS_1 \,,\\
\cA_{25} &= 8U\partial_V \cS_2 ,\\
\cA_{26} &= -8UV[\partial_U+\partial_V] \cS_3 \,.
\end{split}\end{equation}

For $\langle SSSj\rangle$ we find that
\begin{equation}\begin{split}
\cW_{11} &= 2[-1+U(\partial_U+\partial_V)]\cS_2 + 2[1+U(U-1)\partial_U+V\partial_V]\cS_3 \,,\\
\cW_{12} &= 2U\left[U\partial_U+V\partial_V\right]\cS_1+2\left[\frac{2(1-V)} U + (V-1) \partial_U + UV\partial_V\right]\cS_2-\cS_6 \,,\\
\cW_{13} &= 2V\partial_U\cS_1+2\left[\frac{2(1-V)} U + (U+V-1) \partial_U + V\partial_V\right]\cS_3+\cS_5 \,,\\
\cW_{21} &= 2[1+U\partial_U+(U-1)\partial_V]\cS_2 + 2V\left[U\partial_U+V\partial_V\right]\cS_3 + \cS_4  \,,\\
\cW_{22} &= 2UV[\partial_U+(V-1)\partial_V]\cS_1 + 2V\left[\frac 1 u + U^2\partial_U+U\partial_V\right]\cS_2  \,,\\
\cW_{23} &= -2[U\partial_U+(U+V-1)\partial_V]\cS_1+V\left[\frac 2 U - \partial_U\right]\cS_3 - \cS_5  \,.
\end{split}\end{equation}

\section{$\bar D$ functions}
\label{dbarApp}

The quartic Witten contact diagram is given by
 \es{Dfunc2}{
  D_{r_1,r_2,r_3,r_4}(x_i)=\int_{AdS_{d+1}} dz \prod_{i=1}^4G^{r_i}_{B\partial}(z,x_i)\,,\qquad G^{r_i}_{B\partial}(z,x_i)=\left(\frac{z_0}{z_0^2+(\vec z-x_i)^2}\right)^{r_i}\,,
  }
 where $z$ are the $d+1$ bulk spacetime variables and $G^{r_i}_{B\partial}$ is the bulk-to-boundary propagator \cite{Witten:1998qj} for an operator of dimension $r_i$. We can then define the conformally invariant function:
  \es{Dfunc}{
  \bar D_{r_1,r_2,r_3,r_4}(U,V)=\frac{x_{13}^{\frac12\sum_{i=1}^4r_i-r_4}x_{24}^{r_2}}{x_{14}^{\frac12\sum_{i=1}^4r_i-r_1-r_4}x_{34}^{\frac12\sum_{i=1}^4r_i-r_3-r_4}}\frac{2\prod_{i=1}^4\Gamma(r_i)}{\pi^{\frac d2}\Gamma\left(\frac{-d+\sum_{i=1}^4r_i}{2}\right)}D_{r_1,r_2,r_3,r_4}(x_i)\,,
  }
  which is in fact independent of $d$.
  
   The simplest $\bar D_{r_1,r_2,r_3,r_4}(U,V)$ is $\Phi=\bar D_{1,1,1,1}(U,V)$, which is just a scalar one-loop box integral in $d=4$ and can be written as 
  \es{phi}{
  \Phi(z,\bar z)=\frac{1}{z-\bar z}\left(\log(z \bar z)\log\frac{1-z}{1-\bar z}+2\text{Li}(z)-2\text{Li}(\bar z)\right)\,,
  }
  where we define as usual
  \es{zdef}{
  U=z\bar z\,,\qquad V=(1-z)(1-\bar z)\,,
  }
  and $\Phi$ has a recursion relation \cite{Eden:2000bk}
  \es{Phi2}{
  \partial_z\Phi=&-\frac{\Phi}{z-\bar z}-\frac{\log\left((z-1)(\bar z-1)\right)}{z(z-\bar z)}+\frac{\log(z\bar z)}{(z-1)(z-\bar z)}\,,\\
   \partial_{\bar z}\Phi=&\frac{\Phi}{z-\bar z}+\frac{\log\left((z-1)(\bar z-1)\right)}{\bar z(z-\bar z)}-\frac{\log(z\bar z)}{(\bar z-1)(z-\bar z)}\,.
  }
  We can now define the general $\bar D_{r_1,r_2,r_3,r_4}(U,V)$ recursively using the relations \cite{Arutyunov:2002fh}
  \es{Drecurse}{
  \bar D_{r_1+1,r_2+1,r_3,r_4}&=-\partial_U \bar D_{r_1,r_2,r_3,r_4}\,,\\
    \bar D_{r_1,r_2,r_3+1,r_4+1}&=\left(\frac{r_3+r_4-r_1-r_2}{2}-U\partial_U\right) \bar D_{r_1,r_2,r_3,r_4}\,,\\
      \bar D_{r_1,r_2+1,r_3+1,r_4}&=-\partial_V \bar D_{r_1,r_2,r_3,r_4}\,,\\
        \bar D_{r_1+1,r_2,r_3,r_4+1}&= \left(\frac{r_1+r_4-r_2-r_3}{2}-V\partial_V\right) \bar D_{r_1,r_2,r_3,r_4}\,,\\
          \bar D_{r_1,r_2+1,r_3,r_4+1}&=\left(r_2+U\partial_U+V\partial_V\right) \bar D_{r_1,r_2,r_3,r_4}\,,\\
            \bar D_{r_1+1,r_2,r_3+1,r_4}&=\left(\frac{r_1+r_2+r_3-r_4}{2}+V\partial_V+U\partial_U\right) \bar D_{r_1,r_2,r_3,r_4}\,.
  }
  
  Under crossing, the $\bar D_{r_1,r_2,r_3,r_4}$ transform as
  \es{crossDbar}{
  \bar D_{r_1,r_2,r_3,r_4}(V,U)=\bar D_{r_3,r_2,r_1,r_4}(U,V)\,,\quad  \bar D_{r_1,r_2,r_3,r_4}(U^{-1},U/V)= U^{-r_2}\bar D_{r_3,r_2,r_4,r_1}(V,U)\,.
  }

\section{Mellin amplitudes}
\label{MELLIN}

Holographic correlators take a simpler form in Mellin space. To find the Mellin transform of any 4-point function of the form $\langle A_1A_2B_1B_2\rangle$ of scalar operators with scaling dimensions $\Delta_{A_1} = \Delta_{A_2} = \Delta_A$ and $\Delta_{B_1} = \Delta_{B_2} = \Delta_B$, we first define the conformally invariant function $\mathcal{G}^{A_1A_2B_1B_2}(U,V)$ as
\es{Gs}{
\mathcal{G}^{A_1A_2B_1B_2}(U,V)&\equiv x_{12}^{2\Delta_A}x_{34}^{2\Delta_B}\langle A_1(\vec{x}_1)A_2(\vec{x}_2)B_1(\vec{x}_3)B_2(\vec{x}_4)\rangle\,.
}
 We then separate out the disconnected parts of each correlator, which for the correlators we consider take the form
 \es{GDisc}{
  {\cal G}_\text{disc}^{SSSS}(U, V) 
   &= Y_{12}^2Y_{34}^2 + Y_{13}^2Y_{24}^2U+  Y_{14}^2Y_{23}^2\frac{U}{V} \,,\\
     {\cal G}_\text{disc}^{PPPP}(U, V) 
   &= X_{12}^2X_{34}^2 + X_{13}^2X_{24}^2U^2+  X_{14}^2X_{23}^2\frac{U^2}{V^2} \,,\\
     {\cal G}_\text{disc}^{SSPP}(U, V) 
   &= Y_{12}^2X_{34}^2   \,.
 } 
 The Mellin transform $\mathcal{ M}^{A_1A_2B_1B_2}(s,t)$ of the connected part $\mathcal{G}^{A_1A_2B_1B_2}_\text{conn} \equiv {\cal G}^{A_1A_2B_1B_2} - \mathcal{G}_\text{disc}^{A_1A_2B_1B_2}$ is then
 \es{mellin}{
\mathcal{ M}^{A_1A_2B_1B_2}(s,t)=\int^\infty_{-\infty} dUdV V^{-\frac t2+\frac{\Delta_A+\Delta_B}{2}-1}U^{-\frac s2-1}\mathcal{G}^{A_1A_2B_1B_2}_\text{conn}(U,V)\,,
}
which has the inverse transformation
\es{mellinDef}{
\mathcal{G}_\text{conn}^{A_1A_2B_1B_2}(U,V)=\int_{-i\infty}^{i\infty}\frac{ds\, dt}{(4\pi i)^2} U^{\frac s2}V^{\frac t2-\frac{\Delta_A+\Delta_B}{2}}\mathcal{ M}^{A_1A_2B_1B_2}(s,t)\,.
}

For a 4-point function in a large $N$ expansion, it is convenient to consider the auxiliary Mellin amplitude
\es{auxM}{
M^{A_1A_2B_1B_2}(s,t)=\frac{\mathcal{M}^{A_1A_2B_1B_2}(s,t)}{\Gamma\left[\Delta_A-\frac s2\right]\Gamma\left[\Delta_B-\frac s2\right]\Gamma^{2}\left[\frac{\Delta_A+\Delta_B}{2}-\frac t2\right]\Gamma^{2}\left[\frac{\Delta_A+\Delta_B}{2}-\frac u2\right]}\,,
}
where $s+t+u=2\Delta_A+2\Delta_B$, and the Gamma functions automatically encode the pole contribution of all double-trace operators \cite{Mack:2009mi}. The two integration contours in \eqref{mellinDef} then include all poles of the Gamma functions on one side or the other of the contour. 

As an example of the simplicity of holographic correlators in Mellin space, recall that tree level correlators are written in term of $\bar D_{r_1,r_2,r_3,r_4}$ functions, which in position space are given by a complicated recursive algorithm in terms of Dilogarithm functions as described in Appendix \ref{dbarApp}\@. In Mellin space, however, these $\bar D_{r_1,r_2,r_3,r_4}$ contribute to ${ M}^{A_1A_2B_1B_2}(s,t)$ as \cite{Rastelli:2017udc}:
  \es{dbar}{
 & \bar D^{A_1A_2B_1B_2}_{r_1,r_2,r_3,r_4}(s,t)=\left(\Delta_A-\frac{s}{2}\right)_{-\Delta_A}\left(\Delta_B-\frac{s}{2}\right)_{\frac{r_3+r_4-r_1-r_2-2\Delta_B}{2}}\\
&  \left(\frac{\Delta_A+\Delta_B}{2}-\frac{t}{2}\right)_{\frac{r_1+r_4-r_2-r_3}{2}}   \left(\frac{\Delta_A+\Delta_B}{2}-\frac{u}{2}\right)_{r_2}\left(\frac{\Delta_A+\Delta_B}{2}-\frac{u}{2}\right)_{\frac{r_1+r_2+r_3-r_4}{2}}\,,
    }
which for integer $\Delta_A,\Delta_B,r_i$ is a rational function of $s,t,u$. We can get polynomials in $s,t,u$ by shifting $s\to s-2\max\{\Delta_A,\Delta_B\}$, which in position space corresponds to $ U^{\max\{\Delta_A,\Delta_B\}} \bar D_{r_1,r_2,r_3,r_4}$.

\section{Supergravity and $D^4R^4$ terms}
\label{R4D4R4}
The degree 1 Mellin amplitudes for $\langle SSSS\rangle$, $\langle PPPP\rangle$, and $\langle SSPP\rangle$ in the bases \eqref{SbasisM} and \eqref{PbasisM} have the following crossing-independent coefficients
\es{sugraMellinS}{
\tilde \cS^1_1=&-\frac{(t-2) (s+t-2) \left(\sqrt{\pi } (s+4) \Gamma \left(1-\frac{s}{2}\right)-4 \Gamma
   \left(\frac{1-s}{2}\right)\right)}{\sqrt{\pi } s (s+2) \Gamma \left(1-\frac{s}{2}\right)}\,,\\
\tilde \cS^1_4=&\frac{2 (s-2) \left(\frac{\sqrt{\pi } \left(s (t+2)+t^2-4 t-8\right)+\frac{2 t \Gamma \left(\frac{1}{2}
   (s+t-3)\right)}{\Gamma \left(\frac{1}{2} (s+t-2)\right)}}{s+t-4}-\frac{2 \Gamma
   \left(\frac{1-t}{2}\right)}{\Gamma \left(1-\frac{t}{2}\right)}\right)}{\sqrt{\pi } t}\,,
}
and
\es{sugraMellinP}{
\tilde \cP^1_1=&\frac{8 \left(s^2+s (8 t-30)+8 (t-4)^2\right) \Gamma
   \left(\frac{1}{2}-\frac{s}{2}\right)}{\sqrt{\pi }
   (s-2)^2 s (s+2) \Gamma
   \left(1-\frac{s}{2}\right)}\\
   &+\frac{s^3 (42-15 t)+s^2 (3
   (34-5 t) t-176)-2 s (t (9 t-88)+180)+32 (t-4)^2}{s
   \left(s^2-4\right)}\,,\\
\tilde \cP^1_4=&\frac{(s+t-4) \left((124 s-185) t^3+(s (48 s-997)+2474)
   t^2-40 t^4\right) \Gamma \left(\frac{1}{2}
   (s+t-1)\right)}{4 \sqrt{\pi } (t-2)^2 t (s+t-8) (s+t-7)
   (s+t-6) (s+t-5) (s+t-3) \Gamma \left(\frac{1}{2}
   (s+t-2)\right)}\\
   &+\frac{\left((1465-284 s) t^3-2 (s (282
   s-3187)+8663) t^2-40 t^4\right) \Gamma
   \left(\frac{1}{2}-\frac{t}{2}\right)}{8 \sqrt{\pi }
   (t-2)^2 t (s+t-8) (s+t-6)^2 \Gamma
   \left(1-\frac{t}{2}\right)}\\
   &+\frac{4 (s-4)}{t}-\frac{4
   (s-4)}{s+t-8}+\frac{8 (s-3)}{t-2}-\frac{8
   (s-3)}{s+t-6}+30 s-84\\
   &+\frac{(s+t-4) ((s (3 s (4
   s-109)+2896)-6552) t-27 (s-8) (s-6) (s-4)) \Gamma
   \left(\frac{1}{2} (s+t-1)\right)}{4 \sqrt{\pi } (t-2)^2
   t (s+t-8) (s+t-7) (s+t-6) (s+t-5) (s+t-3) \Gamma
   \left(\frac{1}{2} (s+t-2)\right)}\\
   &+\frac{\left(4 (s (2
   (989-56 s) s-11351)+21102) t-128 (s-8) (s-6)^2
   (s-4)\right) \Gamma
   \left(\frac{1}{2}-\frac{t}{2}\right)}{8 \sqrt{\pi }
   (t-2)^2 t (s+t-8) (s+t-6)^2 \Gamma
   \left(1-\frac{t}{2}\right)}\,,
}
and
\es{sugraMellinR}{
\tilde \cR^1_1=&\frac{\left(12 s^2 (24 (t-4) t+73)+16 s (4 t (t (8
   t-63)+146)-367)\right) \Gamma
   \left(\frac{1}{2}-\frac{s}{2}\right)}{32 \sqrt{\pi } s
   (s+2) (t-1) (s+t-5) \Gamma
   \left(2-\frac{s}{2}\right)}\\
   &+\frac{\left(-3 s^4+4 s^3 (8
   t-5)+256 (t-5) (t-3)^2 (t-1)\right) \Gamma
   \left(\frac{1}{2}-\frac{s}{2}\right)}{32 \sqrt{\pi } s
   (s+2) (t-1) (s+t-5) \Gamma
   \left(2-\frac{s}{2}\right)}\\
   &-\frac{3 s^3+4 s^2 (3 t-5)+4
   s (t-2) (3 t-4)+32 (t-3)^2}{4 s (s+2)}\,,\\
\tilde \cR^1_2=&-48 (-2 + s)\,,\\
\tilde \cR^1_3=&\frac{\left(3 s^2+s (26-32 t)-32 (t-5) (t-1)\right) (s+2
   t-6) \Gamma \left(\frac{1}{2}-\frac{s}{2}\right)}{4
   \sqrt{\pi } s (t-1) (s+t-5) \Gamma
   \left(2-\frac{s}{2}\right)}\\
   &+\frac{1}{4}
   \left(\frac{32}{s}+48\right) (s+2 t-6)\,.
}

The degree 6 Mellin amplitudes coefficients are
\es{D4R4MellinS}{
\tilde \cS^6_1=&2 s^4 t^2-\frac{108 s^4 t}{11}+\frac{128 s^4}{11}+6 s^3
   t^3-40 s^3 t^2+\frac{952 s^3 t}{11}-\frac{672
   s^3}{11}+8 s^2 t^4-\frac{752 s^2 t^3}{11}\\
   &+\frac{2280
   s^2 t^2}{11}-\frac{1808 s^2 t}{7}+\frac{8192 s^2}{77}+6
   s t^5-\frac{684 s t^4}{11}+\frac{2624 s
   t^3}{11}-\frac{31760 s t^2}{77}\\
   &+\frac{23328 s
   t}{77}
   -\frac{4736 s}{77}+2 t^6-24 t^5+\frac{1216
   t^4}{11}-\frac{2688 t^3}{11}+\frac{2848
   t^2}{11}-\frac{1152 t}{11}\,,\\
   \tilde \cS^6_4=&-4 s^5 t-\frac{200 s^5}{11}-8 s^4 t^2+\frac{408 s^4
   t}{11}+\frac{2576 s^4}{11}-8 s^3 t^3+\frac{800 s^3
   t^2}{11}-\frac{1296 s^3 t}{11}\\
   &-\frac{90208 s^3}{77}-4
   s^2 t^4+\frac{784 s^2 t^3}{11}-\frac{3056 s^2
   t^2}{11}+\frac{12448 s^2 t}{77}+\frac{219200
   s^2}{77}+\frac{216 s t^4}{11}\\
   &-\frac{2240 s
   t^3}{11}+\frac{40672 s t^2}{77}-\frac{10624 s
   t}{77}-\frac{258816 s}{77}-\frac{256
   t^4}{11}+\frac{2048 t^3}{11}-\frac{30720
   t^2}{77}\\
   &+\frac{8192 t}{77}+\frac{118784}{77}\,,
}
and 
\es{D4R4MellinP}{
\tilde \cP^6_1=&390 s^4 t^2-2236 s^4 t+3232 s^4+1170 s^3 t^3-12896 s^3
   t^2+45384 s^3 t-\frac{571424 s^3}{11}\\
   &+1560 s^2
   t^4-24440 s^2 t^3+142464 s^2 t^2-\frac{3994272 s^2
   t}{11}+\frac{3756736 s^2}{11}+1170 s t^5\\
   &-23140 s
   t^4
   +184176 s t^3-\frac{8080992 s t^2}{11}+1461440 s
   t-\frac{89056256 s}{77}+390 t^6\\
   &-9360 t^5+94168
   t^4-508288 t^3+\frac{17057600 t^2}{11}-\frac{27830784
   t}{11}+\frac{132527104}{77}\,,\\
   \tilde \cP^6_4=&-780 s^5 t-1976 s^5-1560 s^4 t^2+15912 s^4 t+39872
   s^4-1560 s^3 t^3+26624 s^3 t^2\\
   &-133904 s^3
   t-\frac{3575648 s^3}{11}-780 s^2 t^4+21424 s^2
   t^3-184464 s^2 t^2+\frac{6690208 s^2
   t}{11}\\
   &+\frac{14432256 s^2}{11}+4472 s t^4-84480 s
   t^3+\frac{6001696 s t^2}{11}-\frac{15458816 s
   t}{11}-\frac{18638336 s}{7}\\
   &-6464 t^4+103424
   t^3-\frac{6283520 t^2}{11}+\frac{13862912
   t}{11}+\frac{170545152}{77}\,,
}
and
\es{D4R4MellinR}{
\tilde \cR^6_1=&\frac{13 s^5 t}{2}-\frac{27 s^5}{2}+39 s^4 t^2-123 s^4
   t+\frac{978 s^4}{11}+91 s^3 t^3-597 s^3 t^2+\frac{11177
   s^3 t}{11}\\
   &-\frac{3975 s^3}{11}
   +\frac{221 s^2
   t^4}{2}-1104 s^2 t^3+\frac{43117 s^2
   t^2}{11}-\frac{58716 s^2 t}{11}+\frac{316905
   s^2}{154}+78 s t^5\\
   &-1059 s t^4
   +\frac{63880 s
   t^3}{11}-\frac{174162 s t^2}{11}+\frac{1604754 s
   t}{77}-\frac{73041 s}{7}+26 t^6-468 t^5\\
   &+\frac{39266
   t^4}{11}
   -\frac{162312 t^3}{11}+\frac{2669454
   t^2}{77}-\frac{3349188 t}{77}+\frac{1745202}{77}\,,\\
   \tilde \cR^6_2=&416 s^5 t-864 s^5+832 s^4 t^2-8384 s^4 t+13824 s^4+832 s^3
   t^3-12992 s^3 t^2\\
   &+\frac{726208 s^3 t}{11}-\frac{973120
   s^3}{11}+416 s^2 t^4-9216 s^2 t^3+\frac{809280 s^2
   t^2}{11}-\frac{2780928 s^2 t}{11}\\
   &+\frac{3103200
   s^2}{11}-2112 s t^4+30464 s t^3-\frac{1929600 s
   t^2}{11}+\frac{5137152 s t}{11}-\frac{34335296
   s}{77}\\
   &+2560 t^4-30720 t^3+\frac{1618944
   t^2}{11}-\frac{3631104 t}{11}+\frac{1943040}{7}\,,\\
   \tilde \cR^6_3=&-104 s^5 t+216 s^5-416 s^4 t^2+2032 s^4 t-\frac{26832
   s^4}{11}-624 s^3 t^3+4432 s^3 t^2\\
   &-\frac{113360 s^3
   t}{11}+\frac{86000 s^3}{11}-520 s^2 t^4+4224 s^2
   t^3-\frac{122064 s^2 t^2}{11}+\frac{92736 s^2
   t}{11}\\
   &+\frac{200280 s^2}{77}
   -208 s t^5+1280 s
   t^4+\frac{28448 s t^3}{11}-36672 s t^2+\frac{7400112 s
   t}{77}\\
   &-\frac{6253760 s}{77}-736 t^5+11040
   t^4-\frac{731840 t^3}{11}+\frac{2214720
   t^2}{11}-\frac{23554912 t}{77}\\
   &+\frac{14369952}{77}\,.
}

In position space, these degree 6 expressions take the form
 \es{quartic1Pos6}{
 \cS^6_1=&U\left[-\frac{49152}{77}\bar D_{2,1,1,2}+\frac{24576}{7}
  \bar D_{2,1,2,3}-\frac{261120}{77}
  \bar D_{2,1,3,4}+\frac{10240}{11}
  \bar D_{2,1,4,5}\right.\\
  &-\frac{768}{11}
  \bar D_{2,1,5,6}
  +\frac{884736}{77}
  \bar D_{3,1,1,3}-\frac{1730560}{77}
  \bar D_{3,1,2,4}+\frac{135680}{11}
  \bar D_{3,1,3,5}\\
  &-2304\bar D_{3,1,4,6}+128
  \bar D_{3,1,5,7}
  -\frac{250880}{11}
  \bar D_{4,1,1,4}+\frac{224000}{11}
  \bar D_{4,1,2,5}\\
  &-\frac{57600}{11}
  \bar D_{4,1,3,6}+384
  \bar D_{4,1,4,7}+\frac{136960}{11}
  \bar D_{5,1,1,5}-\frac{58368}{11}
  \bar D_{5,1,2,6}+512\bar D_{5,1,3,7}\\
  &\left.-2304
  \bar D_{6,1,1,6}+384\bar D_{6,1,2,7}+128
  \bar D_{7,1,1,7}\right]\,,\\
    \cS^6_4=&U\left[\frac{49152}{77}\bar D_{1,1,2,2}-\frac{866304}{77}
  \bar D_{1,1,3,3}+\frac{1515520}{77}
  \bar D_{1,1,4,4}-\frac{89600}{11}
  \bar D_{1,1,5,5}\right.\\
  &+\frac{9216}{11}
  \bar D_{1,1,6,6}-\frac{276480}{77}
  \bar D_{2,1,2,3}+\frac{1740800}{77}
  \bar D_{2,1,3,4}-\frac{204800}{11}
  \bar D_{2,1,4,5}\\
  &+\frac{46080}{11}
  \bar D_{2,1,5,6}-256
  \bar D_{2,1,6,7}+\frac{307200}{77}
  \bar D_{3,1,2,4}-\frac{158720}{11}
  \bar D_{3,1,3,5}+\frac{64512}{11}
  \bar D_{3,1,4,6}\\
  &-512
  \bar D_{3,1,5,7}-\frac{15360}{11}
  \bar D_{4,1,2,5}+\frac{36864}{11}
  \bar D_{4,1,3,6}-512
  \bar D_{4,1,4,7}+\frac{1536}{11}
  \bar D_{5,1,2,6}\\
  &\left.-256\bar D_{5,1,3,7}\right]\,,
 }
and 
 \es{quartic1Pos6P}{
 \cP^6_1=&U^2\left[\frac{4767744}{77}\bar D_{2,2,2,2}-\frac{25559040}{77}
  \bar D_{2,2,3,3}+\frac{3793920}{11}
  \bar D_{2,2,4,4}-\frac{1136640}{11}
  \bar D_{2,2,5,5}\right.\\
  &+8448
  \bar D_{2,2,6,6}-\frac{25190400}{11}
  \bar D_{3,2,2,3}+\frac{63969280}{11}
  \bar D_{3,2,3,4}-\frac{42024960}{11}
  \bar D_{3,2,4,5}\\
  &+830976\bar D_{3,2,5,6}-53248
  \bar D_{3,2,6,7}+\frac{96424960}{11}
  \bar D_{4,2,2,4}-\frac{126547200}{11}
  \bar D_{4,2,3,5}\\
  &+4455936\bar D_{4,2,4,6}-610688
  \bar D_{4,2,5,7}+24960\bar D_{4,2,6,8}-9242880
  \bar D_{5,2,2,5}\\
  &+6441216\bar D_{5,2,3,6}-1289600
  \bar D_{5,2,4,7}+74880\bar D_{5,2,5,8}+3545088
  \bar D_{6,2,2,6}\\
  &-1256320\bar D_{6,2,3,7}+99840
  \bar D_{6,2,4,8}-524160\bar D_{7,2,2,7}+74880
  \bar D_{7,2,3,8}\\
  &\left.+24960\bar D_{8,2,2,8}\right]\,,\\
    \cP^6_4=&U^2\left[-\frac{442368}{7}\bar D_{2,2,2,2}+\frac{176302080}{77}
  \bar D_{2,2,3,3}-\frac{87664640}{11}
  \bar D_{2,2,4,4}\right.\\
  &+\frac{77690880}{11}
  \bar D_{2,2,5,5}
  -2004480\bar D_{2,2,6,6}+163072
  \bar D_{2,2,7,7}+\frac{4147200}{11}
  \bar D_{3,2,2,3}\\
  &-\frac{70686720}{11}
  \bar D_{3,2,3,4}
  +\frac{129853440}{11}
  \bar D_{3,2,4,5}-5879808\bar D_{3,2,5,6}+988416
  \bar D_{3,2,6,7}\\
  &-49920
  \bar D_{3,2,7,8}
  -\frac{5099520}{11}
  \bar D_{4,2,2,4}+\frac{54328320}{11}
  \bar D_{4,2,3,5}-5437440\bar D_{4,2,4,6}\\
  &+1444352
  \bar D_{4,2,5,7}
  -99840\bar D_{4,2,6,8}+168960
  \bar D_{5,2,2,5}-1311744\bar D_{5,2,3,6}\\
  &+911872
  \bar D_{5,2,4,7}
  -99840\bar D_{5,2,5,8}
  -16896
  \bar D_{6,2,2,6}+106496\bar D_{6,2,3,7}\\
  &\left.-49920
  \bar D_{6,2,4,8}\right]\,,
 }
and
 \es{quartic1Pos6R}{
 \cR^6_1=&U^2\left[\frac{49152}{77}\bar D_{2,2,1,1}-\frac{1142784}{77}
  \bar D_{2,2,2,2}+\frac{2810880}{77}
  \bar D_{2,2,3,3}-\frac{241920}{11}
  \bar D_{2,2,4,4}\right.\\
  &+\frac{44160}{11}
  \bar D_{2,2,5,5}-192
  \bar D_{2,2,6,6}-\frac{3686400}{77}
  \bar D_{3,2,1,2}+\frac{18616320}{77}
  \bar D_{3,2,2,3}\\
  &-\frac{2987520}{11}
  \bar D_{3,2,3,4}+\frac{1085760}{11}
  \bar D_{3,2,4,5}-12288\bar D_{3,2,5,6}+416
  \bar D_{3,2,6,7}\\
  &+\frac{20229120}{77}
  \bar D_{4,2,1,3}-\frac{5802240}{11}
  \bar D_{4,2,2,4}+\frac{3078720}{11}
  \bar D_{4,2,3,5}-49536\bar D_{4,2,4,6}\\
  &+2496
  \bar D_{4,2,5,7}-\frac{3947520}{11}
  \bar D_{5,2,1,4}+\frac{3711360}{11}
  \bar D_{5,2,2,5}-84480\bar D_{5,2,3,6}\\
  &+5824
  \bar D_{5,2,4,7}
  +\frac{1866240}{11}
  \bar D_{6,2,1,5}-74688\bar D_{6,2,2,6}+7072
  \bar D_{6,2,3,7}-29952\bar D_{7,2,1,6}\\
  &\left.+4992
  \bar D_{7,2,2,7}
  +1664\bar D_{8,2,1,7}\right]\,,\\
   \cR^6_2=&U^2\left[-\frac{786432}{77}\bar D_{2,2,2,2}+\frac{1966080}{11}
  \bar D_{2,2,3,3}-\frac{3358720}{11}
  \bar D_{2,2,4,4}+122880\bar D_{2,2,5,5}\right.\\
  &-12288
  \bar D_{2,2,6,6}+\frac{1966080}{11}
  \bar D_{3,2,2,3}-\frac{13926400}{11}
  \bar D_{3,2,3,4}+\frac{13455360}{11}
  \bar D_{3,2,4,5}\\
  &-344064\bar D_{3,2,5,6}+26624
  \bar D_{3,2,6,7}-\frac{3358720}{11}
  \bar D_{4,2,2,4}+\frac{13455360}{11}
  \bar D_{4,2,3,5}\\
  &-516096\bar D_{4,2,4,6}+53248
  \bar D_{4,2,5,7}+122880\bar D_{5,2,2,5}-344064
  \bar D_{5,2,3,6}\\
  &\left.+53248\bar D_{5,2,4,7}-12288
  \bar D_{6,2,2,6}+26624\bar D_{6,2,3,7}\right]\,,\\
   \cR^6_3=&U^2\left[-\frac{393216}{77}\bar D_{2,2,1,1}+\frac{9240576}{77}
  \bar D_{2,2,2,2}-\frac{24207360}{77}
  \bar D_{2,2,3,3}+\frac{2355200}{11}
  \bar D_{2,2,4,4}\right.\\
  &-\frac{522240}{11}
  \bar D_{2,2,5,5}+3072
  \bar D_{2,2,6,6}+\frac{28311552}{77}
  \bar D_{3,2,1,2}-\frac{144752640}{77}
  \bar D_{3,2,2,3}\\
  &+\frac{25026560}{11}
  \bar D_{3,2,3,4}-\frac{10275840}{11}
  \bar D_{3,2,4,5}+141312\bar D_{3,2,5,6}-6656
  \bar D_{3,2,6,7}\\
  &-\frac{18063360}{11}
  \bar D_{4,2,1,3}+\frac{38400000}{11}
  \bar D_{4,2,2,4}-\frac{22625280}{11}
  \bar D_{4,2,3,5}+423936\bar D_{4,2,4,6}\\
  &-26624
  \bar D_{4,2,5,7}+\frac{17530880}{11}
  \bar D_{5,2,1,4}-\frac{19476480}{11}
  \bar D_{5,2,2,5}+503808\bar D_{5,2,3,6}\\
  &-39936
  \bar D_{5,2,4,7}-460800\bar D_{6,2,1,5}+291840
  \bar D_{6,2,2,6}-33280\bar D_{6,2,3,7}\\
  &\left.+36864
  \bar D_{7,2,1,6}-13312\bar D_{7,2,2,7}\right]\,.
    }

\bibliographystyle{ssg}
\bibliography{D4R4}

\end{document}